\documentclass[journal,10pt,twoside]{IEEEtran}

\usepackage{cite}
\usepackage[pdftex]{graphicx}

\usepackage{amsmath}
\usepackage{amssymb}
\usepackage[acronym]{glossaries}
\usepackage{algorithmic}
\usepackage{array}
\usepackage{color}
\usepackage{tikz}
\usetikzlibrary{decorations.pathreplacing}
\usepackage[font=footnotesize]{caption}
\usepackage[font=footnotesize]{subcaption}
\usepackage{dsfont}
\usepackage[ruled]{algorithm2e}
\usepackage{mathtools}

\newacronym{IoT}{IoT}{internet-of-things}
\newacronym{D2D}{D2D}{device-to-device}
\newacronym{UMTS}{UMTS}{Universal Mobile Telecommunications System}
\newacronym{WLAN}{WLAN}{Wireless Local Area Network}
\newacronym{AP}{AP}{access point}
\newacronym{LiFi}{LiFi}{light-fidelity}
\newacronym{FET}{FET}{field-effect transistor}
\newacronym{PED}{PED}{passenger's electronic device}
\newacronym{MCRT}{MCRT}{Monte Carlo ray-tracing}
\newacronym{DL}{DL}{downlink}
\newacronym{UL}{UL}{uplink}
\newacronym{ICI}{ICI}{inter-channel interference}
\newacronym{LED}{LED}{light emitting diode}
\newacronym{PD}{PD}{photo-diode}
\newacronym{IFC}{IFC}{in-flight connectivity}
\newacronym{IFE}{IFE}{in-flight entertainment}
\newacronym{IM/DD}{IM/DD}{intensity-modulation and direct-detection}
\newacronym{IR}{IR}{infra-red}
\newacronym{VL}{VL}{visible light}
\newacronym{UV}{UV}{ultra-violet}
\newacronym{OWC}{OWC}{optical wireless communications}
\newacronym{CAD}{CAD}{computer-aided-design}
\newacronym{VLC}{VLC}{visible light communications}
\newacronym{SIR}{SIR}{signal-to-interference-ratio}
\newacronym{SNR}{SNR}{signal-to-noise-ratio}
\newacronym{SNIR}{SNIR}{signal-to-noise-plus-interference-ratio}
\newacronym{QoS}{QoS}{quality-of-service}
\newacronym{RF}{RF}{radio frequency}
\newacronym{EM}{EM}{electromagnetic}
\newacronym{OOK}{OOK}{on-off keying}
\newacronym{PL}{PL}{path loss}
\newacronym{WDM}{WDM}{wavelength-division-multiplexing}
\newacronym{RGB}{RGB}{red-green-blue}
\newacronym{LoS}{LoS}{line-of-sight}
\newacronym{NLoS}{NLoS}{non-line-of-sight}
\newacronym{VoM}{VoM}{volume of mobility}
\newacronym{FoV}{FoV}{field-of-view}
\newacronym{SM}{SM}{spatial modulation}
\newacronym{OSM}{OSM}{optical spatial modulation}
\newacronym{SISO}{SISO}{single-input-single-output}
\newacronym{MISO}{MISO}{multiple-input-single-output}
\newacronym{MIMO}{MIMO}{multiple-input-multiple-output}
\newacronym{OMIMO}{OMIMO}{optical MIMO}
\newacronym{VR}{VR}{virtual reality}
\newacronym{AR}{AR}{augmented reality}
\newacronym{MTC}{MTC}{machine type communications}
\newacronym{CAGR}{CAGR}{compound annual growth rate}
\newacronym{IM}{IM}{intensity modulation}
\newacronym{DD}{DD}{direct detection}
\newacronym{TX}{TX}{transmitter}
\newacronym{RX}{RX}{receiver}
\newacronym{PAM}{PAM}{pulse amplitude modulation}
\newacronym{MPAM}{$M$-PAM}{$M$-ary pulse amplitude modulation}
\newacronym{QAM}{QAM}{quadrature amplitude modulation}
\newacronym{MQAM}{$M$-QAM}{$M$-ary \gls{QAM}}
\newacronym{PAPR}{PAPR}{peak to average power ratio}
\newacronym{NRZ-OOK}{NRZ-OOK}{non-return-zero on-off keying}
\newacronym{SSK}{SSK}{space shift keying}
\newacronym{GSSK}{GSSK}{generalized \gls{SSK}}
\newacronym{O/E}{O/E}{optical-to-electrical}
\newacronym{E/O}{E/O}{electrical-to-optical}
\newacronym{DAC}{DAC}{digital-to-analog converter}
\newacronym{ADC}{ADC}{analog-to-digital converter}
\newacronym{DC}{DC}{direct-current}
\newacronym{SER}{SER}{symbol error rate}
\newacronym{BER}{BER}{bit error ratio}
\newacronym{OFDM}{OFDM}{orthogonal frequency division multiplexing}
\newacronym{AWGN}{AWGN}{additive white Gaussian noise}
\newacronym{ML}{ML}{maximum-likelihood}
\newacronym{ME-OSM}{ME-OSM}{minimum error OSM}
\newacronym{MU}{MU}{multi user}
\newacronym{bpcu}{bpcu}{bits per channel use}
\newacronym{eU-OFDM}{eU-OFDM}{enhanced unipolar OFDM}
\newacronym{PAM-DMT}{PAM-DMT}{pulse amplitude modulated discrete multitone modulation}
\newacronym{FFT}{FFT}{fast Fourier transform}
\newacronym{IFFT}{IFFT}{inverse \gls{FFT}}
\newacronym{DCT}{DCT}{discrete cosine transform}
\newacronym{IDCT}{IDCT}{inverse \gls{DCT}}
\newacronym{DCO-OFDM}{DCO-OFDM}{direct-current-biased optical \gls{OFDM}}
\newacronym{ACO-OFDM}{ACO-OFDM}{asymmetrically clipped optical \gls{OFDM}}
\newacronym{NDC-OFDM}{NDC-OFDM}{non-DC biased \gls{OFDM}}
\newacronym{PDF}{PDF}{probability density function}
\newacronym{DFT}{DFT}{discrete Fourier transform}
\newacronym{IDFT}{IDFT}{inverse discrete Fourier transform}
\newacronym{ZF}{ZF}{zero-forcing}
\newacronym{MAP}{MAP}{maximum-a-posteriori-probability}
\newacronym{TD-SM}{TD-SM}{time domain \gls{SM}}
\newacronym{FD-SM}{FD-SM}{frequency domain \gls{SM}}
\newacronym{RC}{RC}{repetition coding}
\newacronym{SMX}{SMX}{spatial multiplexing}
\newacronym{STBC}{STBC}{space-time block coding}
\newacronym{SIMO}{SIMO}{single input multiple output}
\newacronym{CCI}{CCI}{co-channel-interference}
\newacronym{OW}{OW}{optical wireless}
\newacronym{PM}{PM}{permutation modulation}
\newacronym{PSK}{PSK}{phase shift keying}
\newacronym{FSK}{FSK}{frequency shift keying}
\newacronym{IMDD}{IM/DD}{intensity modulation and direct detection}
\newacronym{QSM}{QSM}{quadrature \gls{SM}}
\newacronym{GSM}{GSM}{generalized \gls{SM}}
\newacronym{ISI}{ISI}{inter-symbol interference}
\newacronym{CBE-SM}{CBE-SM}{complex-bipolar encoded \gls{SM}}
\newacronym{AH}{AH}{antenna hopping}
\newacronym{CP}{CP}{cyclic prefix}
\newacronym{PMF}{PMF}{probability mass function}
\newacronym{AWG}{AWG}{arbitrary waveform generator}
\newacronym{FBE-SM}{FBE-SM}{fractional bit encoded \gls{SM}}
\newacronym{CSI}{CSI}{channel state information}
\newacronym{STSK}{STSK}{space-time shift keying}
\newacronym{ESM}{ESM}{enhanced \gls{SM}}

\newcommand\figref{Fig.~\ref}
\newcommand{\norm}[1]{ \vert\vert{#1} \rvert\rvert}

\begin{document}

\title{OFDM-Based Optical Spatial Modulation}

\author{Anil~Yesilkaya,~\IEEEmembership{Student~Member,~IEEE,}
        Rui~Bian,~\IEEEmembership{Student~Member,~IEEE,}
        Iman~Tavakkolnia,~\IEEEmembership{Member,~IEEE,}
        and~Harald~Haas,~\IEEEmembership{Fellow,~IEEE}
\thanks{Manuscript received December 31, 2018; revised April 06, 2019; accepted
May 16, 2019. Date of publication XXXX XX, 2019; date of current version
XXXX XX, 2019. The guest editor coordinating the review of this paper and
approving it for publication was Prof. Ananthanarayanan Chockalingam.}
\thanks{Authors are with the LiFi Research and Development Centre, Institute for Digital Communications, School of Engineering, University of Edinburgh, Edinburgh, EH9 3FD, UK. \textit{Anil Yesilkaya, Rui Bian and Iman Tavakkolnia contributed equally to this work.}}}

\markboth{IEEE JOURNAL OF SELECTED TOPICS IN SIGNAL PROCESSING}{Yesilkaya \MakeLowercase{\textit{et al.}}: OFDM-Based Optical Spatial Modulation}

\IEEEoverridecommandlockouts
\IEEEpubid{\makebox[\columnwidth]{\copyright~2019 IEEE. Digital Object Identifier 10.1109/JSTSP.2019.2920577  \hfill} \hspace{\columnsep}\makebox[\columnwidth]{ }}

\maketitle

\IEEEpubidadjcol

\begin{abstract}
\Gls{SM} has proven to be a promising \gls{MIMO} technique which provides high energy efficiency and reduces system complexity. In \gls{SM}, only one transmitter is active at any given time while the rest of them remain silent. The index of the active transmitter carries information. This spatial information is in addition to the data carried by the constellation symbols in the signal domain. Therefore, \gls{SM} increases the transmission rate of the communication system compared to single-input-single-output and \gls{STBC}-\gls{MIMO}. For signal domain data encoding, \gls{OFDM} has been widely adopted. The key benefits in multi-carrier \gls{IM/DD} systems are: i) the capability to achieve high spectral efficiency and ii) the ability to effectively mitigate \gls{DC} wander effects and the impact of ambient light. However, current off-the-shelf \glspl{LED} which are used as transmit entities are primarily bandwidth limited. Thus, there are benefits of combining \gls{SM} and \gls{OFDM} to enhance transmission speeds while maintaining low complexity. In this paper, the two most common \gls{OFDM}-based \gls{SM} types, namely \gls{FD-SM} and \gls{TD-SM}, are investigated for \gls{OWC}. Moreover, proof-of-concept experimental results are presented to showcase practical feasibility of both techniques. The obtained results are also compared with Monte Carlo simulations. The results show that \gls{TD-SM} with an optimal \gls{MAP} detector significantly outperforms \gls{FD-SM}. It can be inferred from the results that \gls{TD-SM} is a strong candidate among \gls{OFDM}-based optical \gls{SM} systems for future optical \gls{IM/DD} wireless communication systems.
\end{abstract}
\glsresetall

\begin{IEEEkeywords}
Spatial modulation (SM), Orthogonal frequency division multiplexing (OFDM), Light fidelity (LiFi), Optical wireless communications (OWC)
\end{IEEEkeywords}

\section{Introduction}
\IEEEPARstart{A}{dvanced} capabilities of mobile devices such as virtual/augmented reality (VR/AR), high definition video/audio streaming, low-latency gaming and \gls{IoT} create a data-greedy ecosystem. The recent forecasts show that the global mobile data traffic will reach 49 exabytes ($\textrm{10}^\textrm{18}$ bytes) by 2021\cite{cisco}. It is also reported in \cite{ch1701} that if the mobile data demand keeps increasing at the same rate as that of the last 10 years, the \gls{RF} spectrum will be completely saturated by 2035. Thus, the higher frequency portion of the spectrum is starting to be considered as a viable solution for the potential congestion. Similar to this trend, \gls{OWC} offers the utilization of the optical portion of the electromagnetic spectrum e.g., \gls{VL}, \gls{IR} and \gls{UV} to create a wireless broadband medium between the ends. Similarly, the \gls{LiFi} which uses \gls{VL} in the downlink and \gls{IR} in the uplink forms a bi-directional optical attocellular network that supports multi-user connectivity and seamless handover\cite{7360112}. In \gls{LiFi}, ordinary off-the-shelf \glspl{LED} and \glspl{PD} will serve as the transmit and receive units, respectively. As the \gls{VL} and \gls{IR} rays are naturally blocked by opaque walls, \gls{LiFi} could achieve very high area spectral efficiency due to significantly reduced co-channel interference in dense network deployments. Moreover, similar features help enhance security and reduce latency in \gls{LiFi}\cite{sbh1301,8302445}. The information is carried by fluctuations in light intensity which is detected by the front-end optics and decoded at the \gls{RX}. This incoherent transmission technique is referred to as \gls{IMDD}. Since the optical power of the \glspl{LED} are intrinsically limited to be real and positive valued, the information bearing signal is also limited by the same fundamental constraints. Furthermore, eye safety and linearity considerations as well as the dynamic range of the front-end opto-electronic elements bring additional limitations to the transmitted signal. Therefore, techniques developed and optimized for \gls{RF} systems cannot be applied straightforwardly to \gls{IMDD} systems without significant changes.

The limited electrical bandwidth of the optical elements motivates the use of \gls{SMX} \cite{to0401,hk0618,b0901,zomfljow0901,dbf1101,fh1301,tsh1301} in \gls{OWC} systems. For a $N_\textrm{t}$ \gls{LED} and $N_\textrm{r}$ \gls{PD} \gls{MIMO}-\gls{OWC} system, the data rate could be enhanced $\min\{N_\textrm{t},N_\textrm{r}\}$-fold by harnessing \gls{SMX}. However, the \gls{ICI} caused by channel coupling and detection complexity at the \gls{RX} side are the major drawbacks of \gls{SMX}. \Gls{SM} is an alternative \gls{MIMO} technique which avoids \gls{ICI} and simplifies the \gls{RX} complexity in exchange for reduced data rates \cite{hcs0201,mhly0501,mhay0601,gmhay0601,mhsay0801,rhgsh1401} compared to ordinary \gls{SMX} systems. In basic \gls{SM} systems, only a single transmitter out of an array of transmitters is \textit{active} at any given time. Therefore, the \gls{ICI} at the receiver side is completely eliminated along with reduced search space at the detector. Furthermore, the channel matrix effectively becomes a column vector which ensures the invertibility of the channel coupling as long as there are no null elements. The active transmitter is random and is selected based on a subset of information bits. The index of the active transmitter carries the \textit{spatial symbol} while the emitted signal from the active transmitter conveys the \textit{constellation symbol} simultaneously. The adoption of \gls{OFDM} in \gls{SM} is proposed for the first time in \cite{gmhay0601}. \gls{OFDM} in conjunction with \gls{SM} applied to optical systems is proposed in \cite{zdsh1201}. In the proposed system, each symbol conveyed by a subcarrier is also accompanied by a spatial symbol such that for each subcarrier there is only one active transmitter. This technique will be referred to as \gls{FD-SM} throughout the paper.  It should be noted that \gls{FD-SM} cannot fully benefit from the advantages of initial \gls{SM} systems. As each \gls{LED} is associated with an \gls{OFDM} stream, $N_\textrm{t}$ modulators and $N_\textrm{r}$ demodulators are required at the \gls{TX} and \gls{RX} sides, respectively. Moreover, \gls{ICI} in \gls{FD-SM} becomes inevitable due to the non-zero time domain samples. Hence, the system performance could become severely degraded if the channel matrix is \textit{ill conditioned}.

In order to address the issues associated with \gls{FD-SM}, an alternative method is proposed in \cite{bel1401} for the optical systems which will be referred to as \gls{TD-SM} throughout the paper. In \gls{TD-SM}, after parallel-to-serial conversion at the \gls{OFDM} modulator, each time domain sample is accompanied by a spatial symbol. Accordingly, each time domain sample carries a spatial symbol along with a constellation symbol. Thus, a single \gls{OFDM} modulator and demodulator pair suffice. Moreover, \gls{ICI} is also completely avoided since only a single transmitter is active per time instant as the \gls{SM} principle suggests. The \gls{TD-SM} technique has been advanced and combined with an optimal \gls{MAP} detector in \cite{tyh1801}.

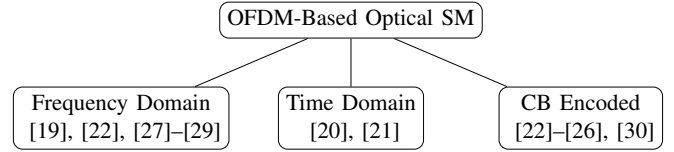
\begin{figure}[!t]
\centering
\resizebox {\columnwidth} {!} {
\begin{tikzpicture}[sibling distance=10em,
  every node/.style = {shape=rectangle, rounded corners,
    draw, align=center,
    top color=white, bottom color=white}]]
  \node {OFDM-Based Optical SM}
    child { node {Frequency Domain\\ \cite{zdsh1201,lth1301,itsymgdvh1401,itmxrcfgdoh1501,hwa1601}} }
    child { node {Time Domain\\ \cite{bel1401,tyh1801}} }
    child { node {CB Encoded\\ \cite{lth1301,bpuh1601,tnc1601,ybmph1701,czw1701,tkka1801}} };
\end{tikzpicture}
}
 \caption{Taxonomy of the \gls{OFDM}-based optical \gls{SM} systems.}
    \label{fig:taxonomy}
\end{figure}

Another unique application of \gls{SM} to \gls{IMDD} systems is the spatial complex-bipolar number encoding. Unlike conventional \gls{SM}, the spatial domain is exploited to realize the transmission of complex and bipolar valued signals in \gls{IMDD} systems. In \cite{lth1301}, a $2\times 2$ system is proposed in which the positive valued real \gls{OFDM} time domain samples are transmitted from the first \gls{LED}. Moreover, the absolute value of the negative samples are also sent simultaneously from the second \gls{LED} in the proposed method. Thus, the constraint of having strictly positive valued transmission signal as imposed by \gls{IMDD} is efficiently overcome. Note that in \cite{lth1301}, the reality of the time domain samples are ensured by applying Hermitian symmetry in the frequency domain. Similarly, another $2\times 2$ technique is given in \cite{czw1701}, suggesting the transmission of the positive valued real and imaginary parts from the first and second \glspl{LED}, respectively. Note that the Hermitian symmetry requirement is removed in \cite{czw1701} by spatial encoding and the positive values are ensured by a \gls{DC} bias and zero clipping. In \cite{bpuh1601,tnc1601,ybmph1701}, both limitations of \gls{IMDD} are jointly compromised by the proposed $4\times 4$ system. Accordingly, positive real and imaginary parts of \gls{OFDM} time domain samples are transmitted from the first and third \glspl{LED} while absolute values of the negative real and imaginary from the second and fourth \glspl{LED}, respectively. Therefore, neither Hermitian symmetry nor \gls{DC} bias are required in \cite{bpuh1601,tnc1601,ybmph1701}. These spatial real/complex and positive/negative encoding systems are referred to as \gls{CBE-SM} throughout the paper. Similar to \gls{FD-SM}, \gls{CBE-SM} cannot also benefit from the advantages of \gls{SM} with increasing $N_\textrm{t}$. For large \gls{MIMO} systems, \gls{CBE-SM} introduces \gls{ICI} and complexity penalties. For convenience and as an overview, \gls{OFDM}-based optical \gls{SM} systems are summarized in \figref{fig:taxonomy}.

In this paper, the system models for both \gls{FD-SM} and \gls{TD-SM} are introduced. A \gls{BER} performance comparison for both systems is conducted using Monte Carlo simulations. Lastly, the comparison is corroborated for the first time by an experimental proof-of-concept study of both FD-SM and TD-SM techniques.

The rest of the paper is organized as follows: a brief history of spatial index modulation is provided in Section II. The channel characteristics for the \gls{OWC} systems are given in Section III. The details of the system models of \gls{FD-SM} and \gls{TD-SM} systems are given in Section IV. In Section V, the experimental setup and methodology are introduced. Computer simulations and experimental results are provided in Section VI. Finally, conclusions are drawn in Section VII.

\emph{Notation}: Throughout the paper, matrices and column vectors are in bold uppercase and bold lowercase letters, respectively. The $m^\textrm{th}$ row and $n^\textrm{th}$ column element of a matrix $\mathbf{A}$ is given by $A(m,n)$. Similarly, the $k^\textrm{th}$ element of a vector $\mathbf{a}$ is denoted by $a(k)$. The transpose of a matrix/vector and the Euclidean norm of a vector are expressed by $(\cdot)^\textrm{T}$ and $\norm{\cdot}$, respectively. A real normal distribution with mean $\mu$ and variance $\sigma^2$ is represented by $\mathcal{N}(\mu,\sigma^2)$. The argument maximum, minimum, argument minimum, complex conjugate, modulus and statistical expectation operators are denoted by $\arg\max\{\cdot\}$, $\min\{\cdot\}$, $\arg\min\{\cdot\}$, $(\cdot)^\ast$, $\vert \cdot \vert$ and $\textrm{E}\{\cdot\}$, respectively. The dot product of two vectors $\mathbf{a}$ and $\mathbf{b}$ is given by $\mathbf{a}\cdot\mathbf{b}$. The continuous time unit step, Dirac delta and Q functions are expressed by $u(\cdot)$, $\delta(\cdot)$ and $\textrm{Q}(\cdot)$, respectively. The Q-function is defined by $\textrm{Q}(x)=\frac{1}{\sqrt{2\pi}}\int_{x}^{\infty}e^{-(u^2/2)}du$. Also, the discrete time Dirac delta function is given by $\delta[\cdot]$. Lastly, the $m\times 1$ zeros vector is denoted by $\mathbf{0}_{m\times 1}$.
\begin{figure*}[!t]
\begin{center}
\begin{tikzpicture}[scale=0.8]
\draw [thick,->] (-11,0) -- (11,0);
\draw [thick,->] (-10,-0.80) -- (-10,-0.15);
\node[align=center] at (-10,-1.7) {1965 \\ Permutation \\ Modulation\cite{s6501}};
\draw [thick,->] (-5,0.80) -- (-5,0.15);
\node[align=center] at (-5,1.7) {1990 \\ Antenna \\ Hopping \cite{b9001}};
\draw [thick,->] (-1.6,-0.80) -- (-1.6,-0.15);
\node[align=center] at (-1.6,-1.7) {2001 \\ Space Shift \\ Keying\cite{cy0101}};
\draw [thick,->] (-0.5,0.80) -- (-0.5,0.15);
\node[align=center] at (-0.5,2) {2002 \\ Spatial Data \\ Multiplexing\\(SDM)\cite{hcs0201}};
\draw [thick,->] (1.0,-0.80) -- (1.0,-0.15);
\node[align=center] at (1.0,-1.7) {2004 \\ AH capacity \\ analysis\cite{syxxjj0401}};
\draw [thick,->] (2.8,0.80) -- (2.8,0.15);
\node[align=center] at (2.8,1.7) {2005 \\ Benefits of \\ SDM \cite{mhly0501}};
\draw [thick,->] (3.9,-0.80) -- (3.9,-0.15);
\node[align=center] at (3.9,-2) {2006 \\ Spatial \\ Modulation\cite{mhay0601} \\/with OFDM \cite{gmhay0601}};
\draw [thick,->] (5.5,0.80) -- (5.5,0.15);
\node[align=center] at (5.5,1.7) {2008 \\ Generalized \\ SSK\cite{jgs0801}};
\draw [thick,->] (7.2,-0.80) -- (7.2,-0.15);
\node[align=center] at (7.2,-2.3) {2010 \\ Generalized \\SM\cite{ysmh1001,fhxyh1001}, \\ FBE-SM\cite{srsmh1001}, \\ DSTSK\cite{sch1001}};
\draw [thick,->] (8.5,0.80) -- (8.5,0.15);
\node[align=center] at (8.5,1.7) {2012 \\ MA-GSM\cite{wjs1201}, \\ GSSK-OWC\cite{pph1201}};
\draw [thick,->] (9.9,-0.80) -- (9.9,-0.15);
\node[align=center] at (9.9,-2.3) {2015 \\ Quadrature \\ SM\cite{mia1501}, \\ Enhanced \\ SM\cite{csss1501}};
\end{tikzpicture}
\end{center}
\caption{History of the SM-based systems.}
\label{fig:smhist}
\end{figure*}
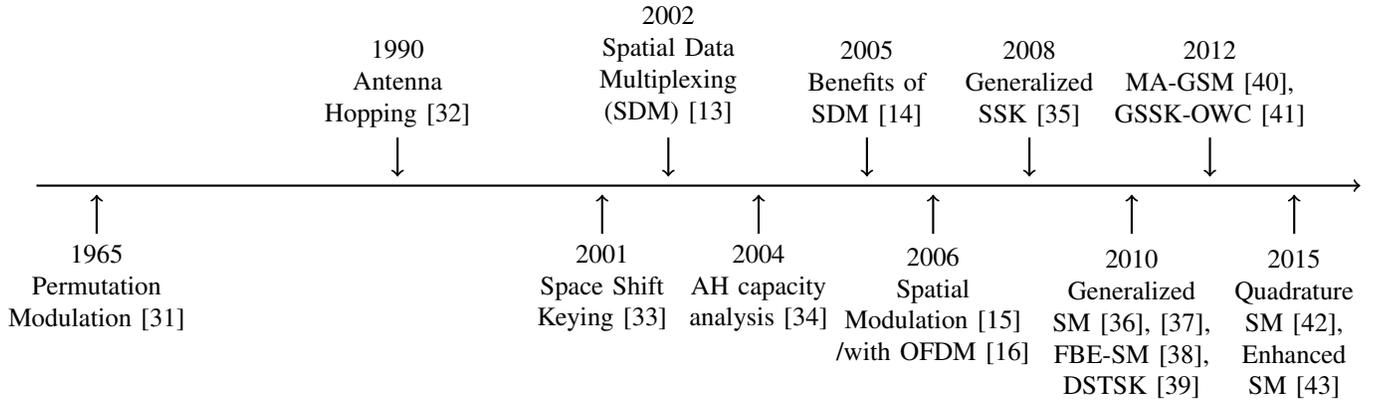

\section{A brief history of spatial index modulation}
A brief summary of spatial index modulation techniques applied to  \gls{RF} and \gls{OWC} systems is given in \figref{fig:smhist}. The very first idea of encoding information to the permutation of the transmit entities is shown by Slepian in 1965 as \gls{PM}\cite{s6501}. \gls{PM} paved the way for \gls{SM} systems which later became a part of the more generalized concept called index modulation. The interested reader may be referred to the following overview paper on \gls{SM}~\cite{rhg1101}. It took 25 years for the first application of \gls{PM} to appear in the spatial domain, referred to as \gls{AH}, which is proposed in \cite{b9001}. The capacity analysis for the channel/antenna hopping technique is carried out in \cite{syxxjj0401}. In 2001, a space modulation technique referred to as \gls{SSK} which uses different channel conditions to encode and decode the information is proposed \cite{cy0101}. Unlike \gls{SM}, \gls{SSK} only encodes information in spatial symbols, and it does not encode information in the signal space. As a consequence, \gls{SSK} exhibits an additional complexity reduction, but the achievable spectral efficiencies are limited. \gls{SSK} is exhaustively analyzed for \gls{RF} communication systems under different channel conditions \cite{rh1001,rh1002}. A novel data multiplexing  technique is proposed in \cite{hcs0201} for a $4\times4$ system that carries all the information of the four streams by activating a single antenna per transmission step. However, the system required parity bits which compromised spectrum efficiency. The work in \cite{hcs0201} has been extended in \cite{mhly0501} where the limitations of parity bits is overcome completely. The new technique was referred to as \gls{SM} for the first time in \cite{mhay0601}. In 2006, an \gls{OFDM}-based \gls{SM} is also proposed\cite{gmhay0601}. The generalization of \gls{SSK}, namely \gls{GSSK} is proposed in 2008\cite{jgs0801}. The number of active antennas in \gls{GSSK} could be arbitrarily chosen which increases the spectral efficiency significantly. In a similar manner, the generalization of \gls{SM}, namely \gls{GSM} is proposed in 2010 \cite{ysmh1001,fhxyh1001}. It is worth noting that the active transmitters in \gls{GSM} transmit the same information in order to avoid \gls{ICI}. In the same year, the first application of \gls{SM} to \gls{OWC} systems is also presented\cite{mmeh1001}. Furthermore, a novel technique to allow arbitrary number of transmitters in \gls{SM} systems namely \gls{FBE-SM} is also proposed\cite{srsmh1001}. Another novel method which constitutes the framework of the differential \gls{SM} systems, namely differential \gls{STSK} is proposed in \cite{sch1001}. In differential \gls{STSK}, the necessity for the receiver side channel state information is removed. In \cite{wjs1201} and \cite{nrnv1301}, another generalization of \gls{SM} is proposed, such that each active antenna transmits different signal in exchange for \gls{ICI} and increased detection complexity. The generalization of \gls{SSK} for \gls{OWC} by relaxation of the number of active \glspl{LED} is proposed in \cite{pph1201}. Accordingly, the number of active \glspl{LED} could take any value per transmitted symbol which increases the spectral efficiency significantly. In 2015, a technique, named \gls{QSM} to enhance the spectral efficiency of \gls{SM} is proposed\cite{mia1501}. In \gls{QSM}, the I (in-phase) and Q (quadrature) parts of a complex bipolar constellation symbol are separated and transmitted from antennas chosen independently. It is important to note that the chosen antenna for I and Q could coincide, which will not introduce \gls{ICI} as both parts are orthogonal. As both I and Q parts carry different spatial symbols, the total data carried in \gls{QSM} becomes larger compared to conventional \gls{SM}. Unlike conventional \gls{SM} methods, two antennas are active per given time instant  in \gls{QSM} unless I and Q parts of the symbol concur. This basic idea is used to enhance the number of transmit symbol possibilities in a technique referred to as \gls{ESM}\cite{csss1501}. In \gls{ESM}, two different constellation sets, $M$-ary and $\frac{M}{2}$-ary, are employed as primary and secondary modulations. Hence, $M$-ary constellation is employed if the number of active antennas is one where $\frac{M}{2}$-ary constellation is used when there are two active antennas. It should also be noted that the number of the transmit combinations in both \gls{QSM} and \gls{ESM} are the same, though \gls{ESM} has more flexibility on the transmit constellation design. 
\section{OWC Channel Model}
\begin{figure}[!b]
	\centering
	\includegraphics[width=0.9\columnwidth]{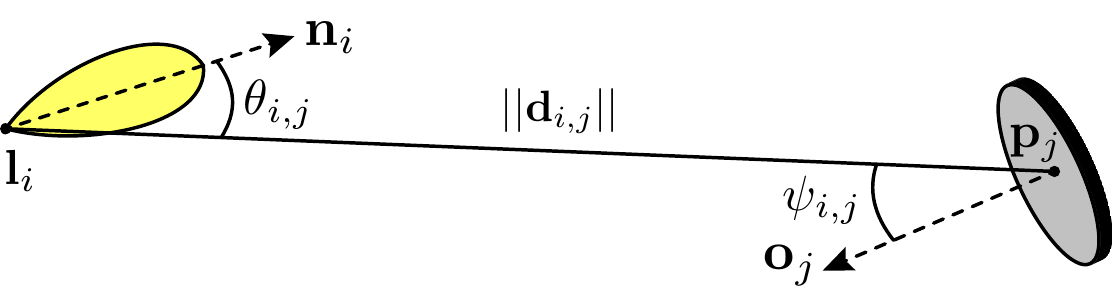}
	\caption{Geometrical model of the point-to-point optical wireless link}
	\label{fig:channel}
\end{figure}
\begin{figure*}[!t]
        \centering
            \centering
            \includegraphics[width=0.9\textwidth,draft=false]{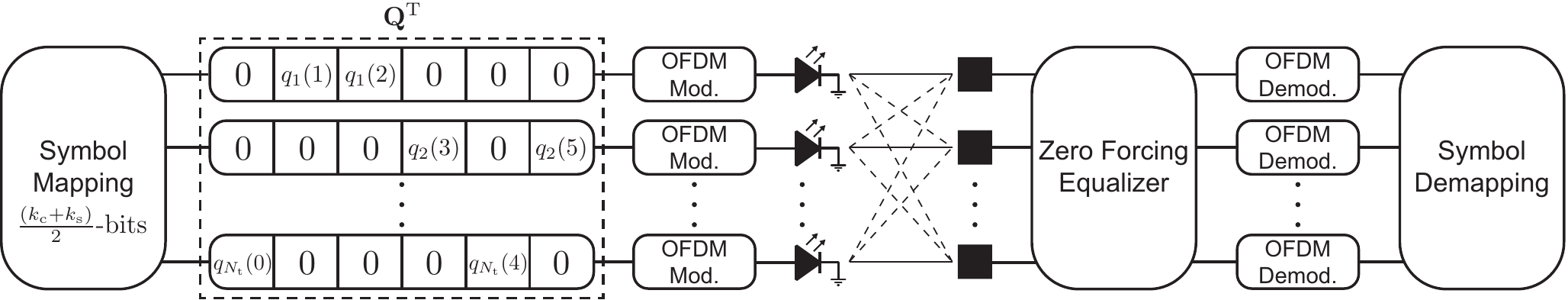}
            \caption{The block diagram of the FD-SM system.}
\label{fig:FDSM}
\end{figure*}
In this section, the properties of the optical wireless channel are presented. Throughout the paper, a $N_\textrm{t}\times N_\textrm{r}$ non-imaging receiver based point-to-point \gls{MIMO}-\gls{OWC} link will be considered. It is reported in \cite{cbh1601} that the multipath effects arising from the side walls, ceiling and floor reflections are negligible as long as a dominant \gls{LoS} link exists. Thus, only the \gls{LoS} \gls{DC} channel gains are considered in this work. It is worth noting that all the \glspl{LED} are assumed to be operating within their linear dynamic range. The geometric parameters of the \gls{OWC} link are depicted in \figref{fig:channel}. Accordingly, the position vectors of the $i^\textrm{th}$ \gls{LED} and $j^\textrm{th}$ \gls{PD} are given by $\mathbf{l}_i$ and $\mathbf{p}_j$, respectively. Moreover, the normal vectors of the $i^\textrm{th}$ \gls{LED} and $j^\textrm{th}$ \gls{PD} are denoted by $\mathbf{n}_i$ and $\mathbf{o}_j$, respectively. The \gls{LoS} \gls{DC} optical wireless channel gains between $i^\textrm{th}$ \gls{LED} and $j^\textrm{th}$ \gls{PD} are given as follows\cite{kb9701}:
\begin{equation}
H(j,i)=\frac{(m+1)A_{\textrm{PD}}}{2\pi \norm{\mathbf{d}_{j,i}}^2} \cos^m(\theta_{j,i})\cos(\psi_{j,i})\mathds{1}_{\kappa_{1/2}}\left(\psi_{j,i} \right),
\label{eq:channel}\end{equation}
\noindent where $i\in\{1,2,\cdots,N_\textrm{t}\}$ and $j\in\{1,2,\cdots,N_\textrm{r}\}$. The parameter $m$ is the Lambertian mode number which is defined by \mbox{$m=-1/\log_2(\cos(\Phi_{1/2}))$}. The semi-angle of the half power of the \glspl{LED} is denoted by $\Phi_{1/2}$. Note that all the \glspl{LED} and \glspl{PD} are assumed to be identical. The detection area of the \glspl{PD} is given by $A_{\textrm{PD}}$. The radiance and incidence angles between $i^\textrm{th}$ \gls{LED} and $j^\textrm{th}$ \gls{PD} are denoted by $\theta_{j,i}$ and $\psi_{j,i}$, respectively. The Euclidean distance vector from $j^\textrm{th}$ \gls{PD} to $i^\textrm{th}$ \gls{LED} is given by
\begin{equation*}
\mathbf{d}_{j,i}=\mathbf{l}_i-\mathbf{p}_j.
\end{equation*}
\noindent The radiance and incidence angles could alternatively be represented as follows:
\begin{equation}
\theta_{j,i}=\arccos\frac{-\mathbf{n}_i\cdot \mathbf{d}_{j,i}}{\norm{\mathbf{d}_{j,i}}} ~~\textrm{and}~~ \psi_{j,i}=\arccos\frac{\mathbf{o}_j\cdot \mathbf{d}_{j,i}}{\norm{\mathbf{d}_{j,i}}}.
\label{eq:channel2}\end{equation}
\noindent By combining \eqref{eq:channel} and \eqref{eq:channel2}, we can express the \gls{DC} channel gains by
\begin{equation}
H(j,i)=\frac{A_\textrm{PD}(m+1)(-\mathbf{n}_i\cdot\mathbf{d}_{j,i})^m(\mathbf{o}_j\cdot \mathbf{d}_{j,i})}{2\pi \norm{\mathbf{d}_{j,i}}^{m+3}}\mathds{1}_{\kappa_{1/2}}\left(\psi_{j,i} \right).
\end{equation}
\noindent The indicator function $\mathds{1}_{\kappa_{1/2}}\left(\cdot \right)$, determines whether the incident rays are in the \gls{FoV} of the \gls{PD} or not
\[
    \mathds{1}_{\kappa_{1/2}}\left(x \right) = \left\{\begin{array}{lr}
        1, & \text{if } \vert x \vert\leq \kappa_{1/2}\\
        0, & \text{otherwise}.
        \end{array}\right.
 \]
\noindent The \gls{FoV} semiangle of the \gls{PD} is denoted by $ \kappa_{1/2}$. Consequently, the $N_\textrm{r}\times N_\textrm{t}$ channel matrix could be obtained as $\mathbf{H}=H(j,i),~\forall i,j$.

Like in any communication system, the performance of  \gls{OWC} systems is heavily affected by noise. However, in \gls{OWC}, there are two major types of noise sources, namely i) shot noise and ii) thermal noise. The photo-generated shot noise occurs due to the discrete excitement levels in the photo-detector devices. For an \gls{OWC} link, the shot noise, which generally follows a Poisson distribution, arises due to the transmitted signal and ambient light. Unlike shot noise, thermal noise is due to the front-end circuitry and thus signal independent. Thermal noise could be modeled by a Gaussian distribution\cite{kn0301}. In practical \gls{LiFi} systems, the \gls{DC} forward current of \glspl{LED} are relatively high compared to the information carrying signal in order to meet the illumination requirements. Hence, the overall noise is dominated by the photo-induced shot noise. Moreover, the high intensity shot noise consists of many independent filtered Poisson processes. Under these circumstances the central limit theorem can be applied and the total noise  typically follows a Gaussian distribution\cite{hranilovic}. Consequently, the effective noise in \gls{OWC} systems shows \gls{AWGN} characteristics.

\section{OFDM-Based Optical Spatial Modulation}
In this section, we provide a deeper insight into \gls{OFDM}-based \gls{SM} for \gls{MIMO}-\gls{OWC} systems. Then, the \gls{FD-SM} approach and the recently proposed \gls{TD-SM} technique will be explained in greater detail.
\begin{figure*}[!t]
        \centering
            \centering
            \includegraphics[width=0.9\textwidth,draft=false]{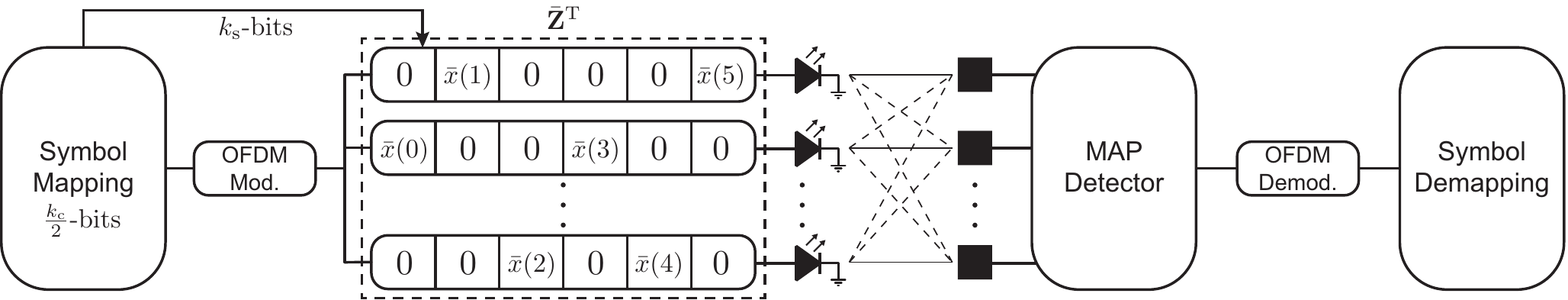}
            \caption{The block diagram of the TD-SM system.}
\label{fig:TDSM}
\end{figure*}
\subsection{Conventional Frequency Domain SM}
\gls{FD-SM}  is a widely adopted \gls{OFDM}-based \gls{SM} technique used in both \gls{RF} and optical domains. The block diagram of the $N_\textrm{t} \times N_\textrm{r}$ \gls{FD-SM} system is given in \figref{fig:FDSM}. The incoming user bits are partitioned into $(k_\textrm{c}+k_\textrm{s})$-bits where the number of constellation and spatial bits are given by $k_\textrm{c}=\log_2(M)$ and $k_\textrm{s}=\log_2(N_\textrm{t})$, respectively. The constellation modulation order is denoted  by $M$. The following one-to-one mapping encodes locations and values of active subcarriers to a $\left(\frac{N}{2}-1\right) \times N_\textrm{t}$ subcarriers matrix $\mathbf{Q}$. The parameter $N$ denotes the \gls{FFT} size. Accordingly, the columns of $\mathbf{Q}$ represent independent \gls{OFDM} streams, which will each be associated with an \gls{LED}. The rows of $\mathbf{Q}$ indicate the subcarrier level spatial information. In other words, the index and value of an active subcarrier convey the spatial and constellation symbols, respectively. The active subcarriers are modulated by \gls{MQAM}. Following the \gls{SM} principle, only a single subcarrier has a non-zero value per row of $\mathbf{Q}$. The $i^\textrm{th}$ column vector of $\mathbf{Q}$ is denoted by $\mathbf{q}_i$. The \gls{PMF} of the independently identically distributed (i.i.d.) elements of $\mathbf{q}_i$ is given by
\begin{equation}
p_{\mathbf{q}_i}[w]=\frac{1+\left( M(N_\textrm{t}-1)-1\right)\delta[\vert w \vert]}{M N_\textrm{t}},
\end{equation}
\noindent for $w\in \{ 0,\mathcal{C} \}$. The \gls{MQAM} constellation alphabet is denoted by $\mathcal{C}$. Thus, the electrical power of $\mathbf{q}_i$ becomes $\textrm{E}\{ \mathbf{q}_i^2\}=1/N_\textrm{t}$. Note that the average electrical power of the \gls{MQAM} constellation symbols are set to unity. In order to ensure that the resultant signal is real after the \gls{IFFT} operation, Hermitian symmetry is imposed on $\mathbf{Q}$ as follows:
\begin{equation}
\mathbf{S}=[\mathbf{0}_{N_\textrm{t}\times 1}~~\mathbf{Q}^\textrm{T}~~\mathbf{0}_{N_\textrm{t}\times 1}~~\tilde{\mathbf{Q}}^\textrm{T}]^\textrm{T}.
\end{equation}
The conjugate symmetric matrix of $\mathbf{Q}$ is given by $\tilde{\mathbf{Q}}$. Therefore, the elements of $\mathbf{S}$ become, $S(k,i)=S(N-k,i)^\ast,~\forall i,$ where $k\in \{ 0,1,\cdots,N-1\}$ and $S(0,i)=S(N/2,i)=0,~\forall i$. Then, the real and bipolar time domain sample vectors are obtained by applying an $N$-point \gls{IFFT} to the $i^\textrm{th}$ column vector of $\mathbf{S}$,
\begin{equation}
\mathbf{x}_i=\mathbf{F}^{-1}\mathbf{s}_i,\quad \textrm{where }i\in \{ 1,2,\cdots,N_\textrm{t}\}.
\end{equation}
\noindent The $N\times N$ \gls{IFFT} matrix is given by $\mathbf{F}^{-1}$. Moreover, the $i^\textrm{th}$ column vector of the resultant $N \times N_\textrm{t}$ time domain samples matrix $\mathbf{X}$ is denoted by $\mathbf{x}_i$. For a relatively large \gls{IFFT} size ($N>64$), the real and bipolar i.i.d. $\mathbf{x}_i$ vectors are approximated with a Gaussian distribution, $\mathbf{x}_i\sim \mathcal{N}(0,\sigma^2)$ where $\sigma^2=\frac{1}{NN_\textrm{t}}$. In order to satisfy the unipolarity restriction of \gls{IMDD} systems, a \gls{DC} bias is introduced to $\mathbf{x}_i$ and the remaining negative samples are clipped to zero. The element-wise biasing and clipping effects are given by $\bar{\mathbf{x}}_i=\left[ \mathbf{x}_i +B \right]^{U}_{L}$, where $[x]^U_L\coloneqq L+\left(  \min \{x,U\}-L \right)u(x-L)$. The lower and upper clipping values are denoted by $L$ and $U$, respectively. The \gls{DC} bias value is defined by $B=r\sigma$ where $r$ is the bias proportionality constant. The \gls{DC} bias level in decibels is defined by $B_\textrm{dB}=10\log_{10}(r^2+1)$. The \gls{PDF} of the biased and clipped elements of the $\bar{\mathbf{x}}_i$ is given by 
\begin{equation}
\begin{split}
f_{\bar{\mathbf{x}}_i}(v)&=\frac{1}{\sqrt{2\pi \sigma^2}}e^{-(v-B)^2/2\sigma^2}\Pi\left( v \right)\\
&+\textrm{Q}\left( \frac{B-L}{\sigma} \right)\delta(v-L)+\textrm{Q}\left(\frac{U-B}{\sigma} \right)\delta(v-U),
\end{split}
\label{eq:pdftime}\end{equation}
\noindent where $\Pi(v)=u(v-L)-u(v-U)$. Note from \eqref{eq:pdftime} that the locations of the lower clipped samples are completely random in $\bar{\mathbf{X}}=[\bar{\mathbf{x}}_1~~\bar{\mathbf{x}}_2~~\cdots~~\bar{\mathbf{x}}_{N_\textrm{t}}]$. Even for the best case scenario, where $L=0$, the likelihood that a single element in $\bar{\mathbf{X}}$ per row will become non-zero is extremely low. Therefore, \gls{ICI} is unavoidable in \gls{FD-SM}. The average transmitted electrical power per \gls{LED} for the \gls{FD-SM} is calculated by using \eqref{eq:pdftime} as follows:
\begin{equation}
\begin{split}
\textrm{E}\{\bar{\mathbf{x}}_i^2\}&=\frac{\sigma}{\sqrt{2\pi}}\left( (L+B)e^{\frac{-(L-B)^2}{2\sigma^2}} - (U+B)e^{\frac{-(U-B)^2}{2\sigma^2}}\right)\\
&+(B^2-L^2+\sigma^2)\textrm{Q}\left( \frac{L-B}{\sigma}\right)\\
&+(U^2-B^2-\sigma^2)\textrm{Q}\left( \frac{U-B}{\sigma}\right)+L^2.
\end{split}
\label{eq:FDSMpower}\end{equation}
\noindent Therefore, the total transmit electrical power of \gls{FD-SM} becomes, $P_{\textrm{e,FD-SM}}=N_\textrm{t}\textrm{E}\{\bar{\mathbf{x}}_i^2\}$. After the \gls{IFFT}, a \gls{CP} sequence of length $N_\textrm{CP}$ is added to each $\bar{\mathbf{x}}_i$. Lastly, digital-to-analog converted $(N+N_\textrm{CP})$-length column vectors are fed to the $N_\textrm{t}$ \glspl{LED} in a serial fashion. The spectral efficiency of \gls{FD-SM} could be determined by using time ($G_\textrm{T}$) and frequency domain ($G_\textrm{F}$) utilization factors as follows:
\begin{equation}
\begin{split}
\eta_{\textrm{FD-SM}}&=\frac{1}{2}\left( k_\textrm{c}+k_\textrm{s} \right)G_\textrm{F}G_\textrm{T}\\
&=\frac{1}{2}\left( k_\textrm{c}+k_\textrm{s} \right)\left( \frac{N-2}{N}\right) \left(\frac{N}{N+N_{\textrm{CP}}}\right)\\
&\approx \frac{1}{2}\left( k_\textrm{c}+k_\textrm{s} \right) \quad\textrm{bits per channel use (bpcu)}.
\end{split}
\label{eq:FDSMspectral}\end{equation}
\noindent The factor $1/2$ in \eqref{eq:FDSMspectral} stems from the Hermitian symmetry imposed on $\mathbf{S}$.

At the receiver side, the optical signals are converted back to the electrical domain by the \glspl{PD}. The $N_\textrm{r}\times N$ real and positive valued baseband received signal matrix after the \gls{CP} removal and analog-to-digital conversion is given by
\begin{equation}
\mathbf{Y}=\mathbf{H}\bar{\mathbf{X}}^\textrm{T}+\mathbf{N},
\end{equation}
\noindent where $\mathbf{N}$ denotes the $N_\textrm{r} \times N$ \gls{AWGN} matrix. The $j^\textrm{th}$ row vector of $\mathbf{N}$ follows an i.i.d Gaussian distribution, \mbox{$\mathbf{n}_j\sim\mathcal{N}(0,\sigma_\textrm{N}^2)$} for $j\in \{1,2,\cdots,N_\textrm{r}\}$. The $N_\textrm{r}\times N_\textrm{t}$ optical channel impulse response matrix, detailed in the previous section, is denoted by $\mathbf{H}$. For simplicity, only the flat frequency response channels are considered in both simulations and experiments throughout the paper. The process of obtaining flat channels in experimental set-up will be detailed in the Section V. Note that \gls{OFDM} would still outperform the single carrier systems due to its inherent resilience to \gls{DC} wander effects and slow ambient light fluctuations (below the minimum modulation frequency which is typically around 1~MHz). The digital-to-analog, electrical-to-optical, optical-to-electrical and analog-to-digital conversion coefficients are taken as unity without loss of generality.

In order to detect the transmitted symbols, the channel coupling effects in the received signal needs to be removed first. Therefore, the channel decoupled received signal matrix which is obtained after the feed-forward equalization is given by $\hat{\mathbf{Y}}=\mathbf{W}\mathbf{Y}$. For the sake of simplicity, the \gls{ZF} equalizer, which is widely adopted in the literature, is chosen as the feed-forward equalizer, $\mathbf{W}=\mathbf{H}^{-1}$. It should be noted that the channel matrix $\mathbf{H}$ becomes singular if the \gls{OWC} link geometry has perfect symmetry. In other words, either rows and/or columns of $\mathbf{H}$ become linearly dependent. In practice, the perfect symmetry of the link is impossible to maintain due to the non-ideal geometry and characteristics of the front-end opto-electronics. However, the similarity between the rows and/or columns of the channel matrix would bring significant performance degradation. Hence, non-singular channel matrices with various condition numbers are considered throughout this paper without loss of generality. The \gls{ZF} equalized and serial-to-parallel converted $N\times N_\textrm{r}$ received time domain samples matrix is given by $\hat{\mathbf{Y}}=\bar{\mathbf{X}}+\mathbf{N}^\textrm{T}\left(\mathbf{H}^\textrm{T}\right)^{-1}$. The $N$-point \gls{FFT} of the $i^\textrm{th}$ column vector of $\hat{\mathbf{Y}}$ is obtained by $\hat{\mathbf{s}}_i=\mathbf{F}\hat{\mathbf{y}}_i$ for $i\in \{1,2,\cdots,N_\textrm{r}\}$. After the \gls{FFT}, the \gls{ZF} equalized subcarriers matrix becomes $\hat{\mathbf{S}}=[\hat{\mathbf{s}}_1~~ \hat{\mathbf{s}}_2~~ \cdots ~~ \hat{\mathbf{s}}_{N_\textrm{r}}]$. Then, the data carrying active subcarrier indexes (spatial symbols) and their values (constellation symbols) are extracted from the $\hat{\mathbf{S}}$. For the optimal detection, both spatial and constellation symbols must be decoded simultaneously\cite{jgs0801}. The joint \gls{ML} detector for the spatial and constellation symbols is given as follows:
\begin{equation}
\begin{split}
\left(\hat{j}(i),\hat{c}(i)\right)&=\arg\max\limits_{\substack{1 \leq j \leq N_\textrm{r} \\ \forall c \in \mathcal{C} }}p\left(\hat{S}(i,j) ~\vert~ c\right)\\
&=\arg\min\limits_{\substack{1 \leq j \leq N_\textrm{r} \\ \forall c \in \mathcal{C} }}\vert \vert \hat{S}(i,j) - c\vert \vert, \\ &\quad \forall i\in \{0,1,\cdots,N-1\}.
\end{split}\label{eq:ML}
\end{equation}
\noindent In \eqref{eq:ML}, the constellation symbol drawn from the \gls{MQAM} set is denoted by $c \in \mathcal{C}$. Moreover, the estimated spatial and constellation symbols are stored for each subcarrier in \mbox{$\hat{\mathbf{j}}=[\hat{j}(0),\hat{j}(1),\cdots,\hat{j}(N-1)]^\textrm{T}$} and \mbox{$\hat{\mathbf{c}}=[\hat{c}(0),\hat{c}(1),\cdots,\hat{c}(N-1)]^\textrm{T}$}, respectively. Finally, the transmitted user bits are simply obtained back by inputting $\hat{\mathbf{j}}$ and $\hat{\mathbf{c}}$ into the inverse mapping function.
\begin{figure}[!t]
        \centering
        \begin{subfigure}[t]{0.8\columnwidth}
            \centering
            \includegraphics[width=1\columnwidth,draft=false]{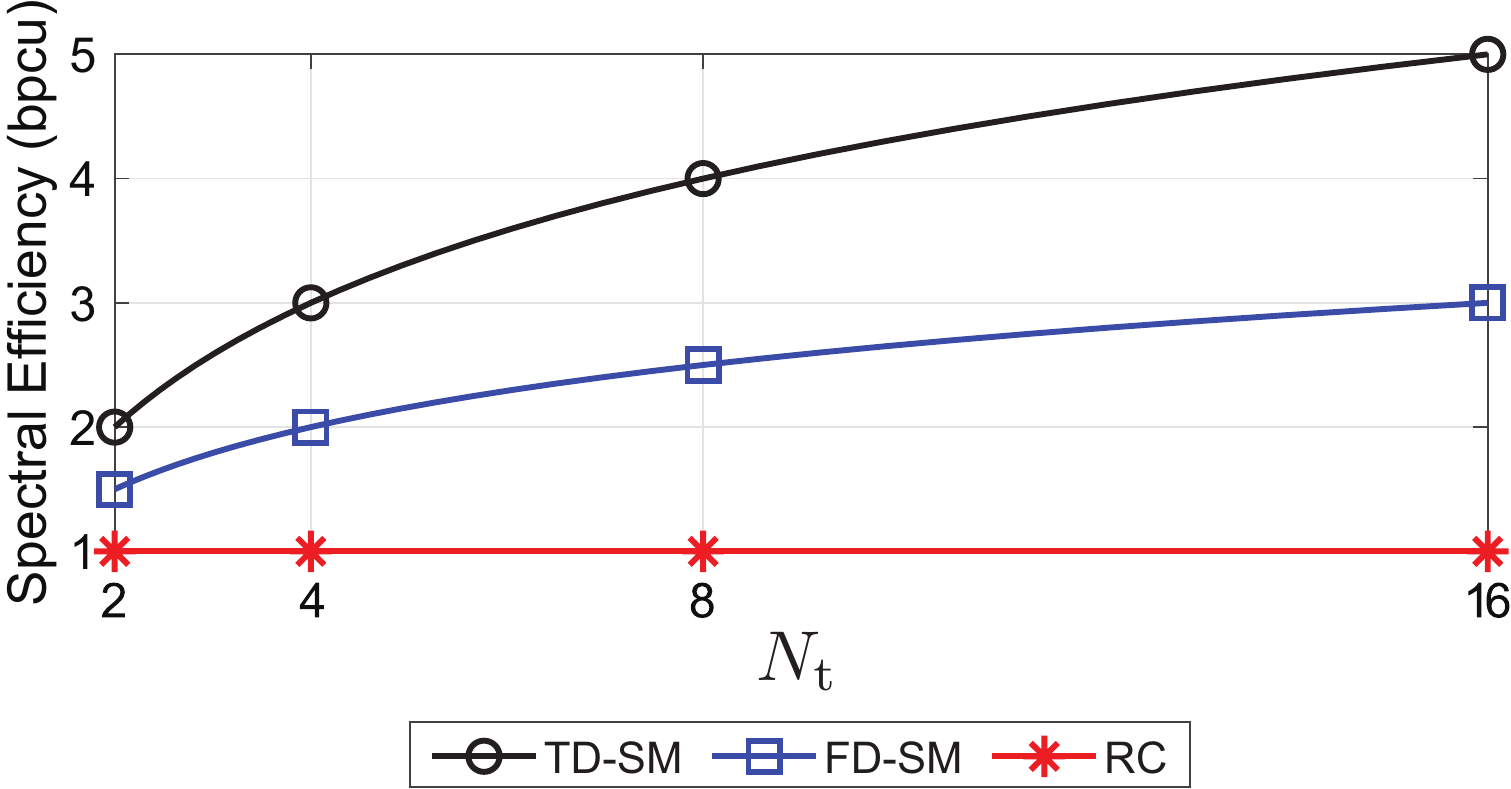}
            \caption{$M=4$}
        \end{subfigure}\\
        \begin{subfigure}[t]{0.8\columnwidth}
            \centering
            \includegraphics[width=1\columnwidth,draft=false]{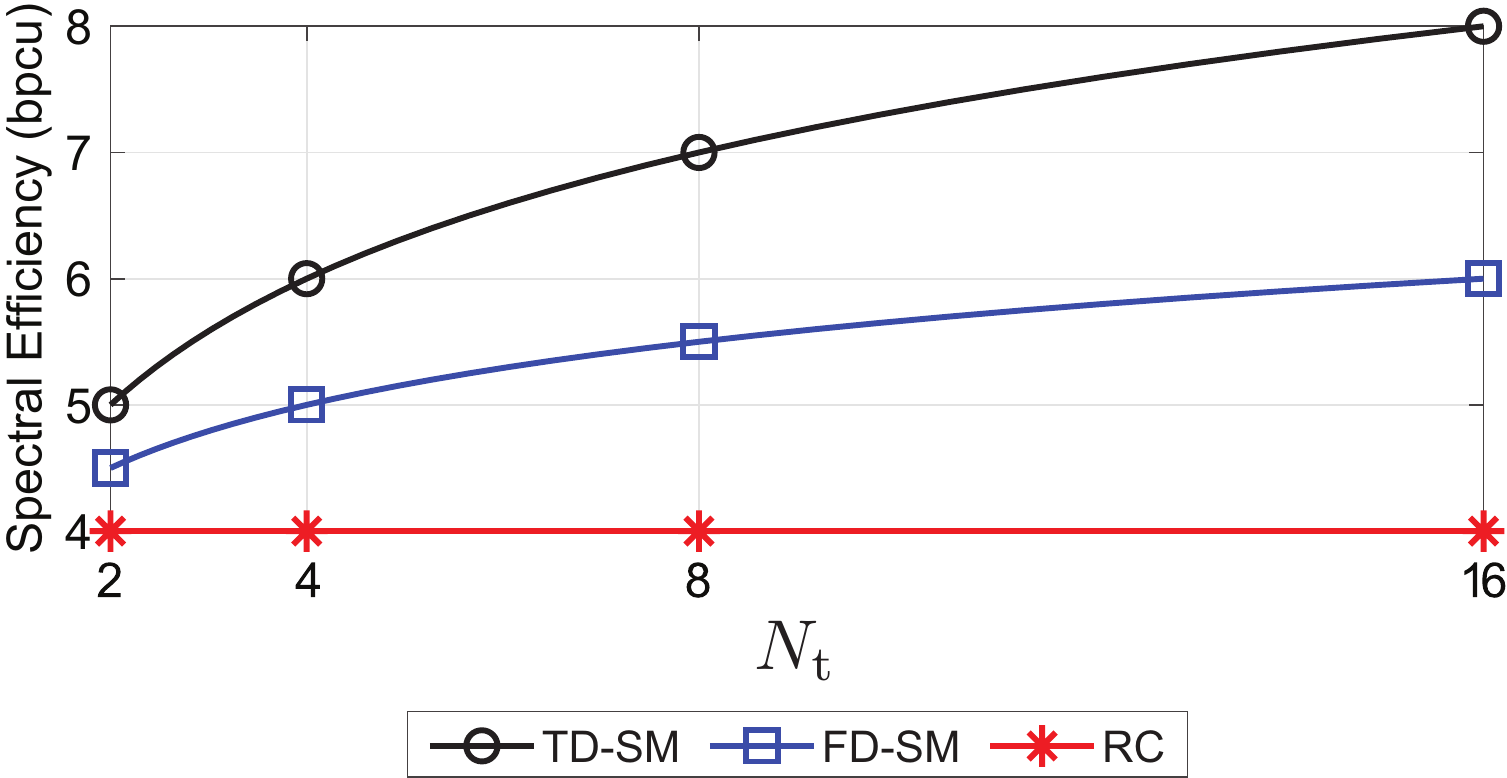}
            \caption{$M=256$}
        \end{subfigure}
        \caption{Spectral efficiency comparison for TD-SM, FD-SM and RC techniques.}
        \label{fig:secomparison}
    \end{figure}
\subsection{Time Domain SM} \label{sec_TDSM}
The fundamental difference between \gls{FD-SM} and \gls{TD-SM} is that \gls{FD-SM} dismisses the main advantages of \gls{SM}  while \gls{TD-SM} retains all of them at the cost of stringent time synchronization requirement. Furthermore, the achievable spectral efficiency in \gls{TD-SM} is significantly higher compared to \gls{FD-SM} due to the time domain spatial symbol mapping. The block diagram of  \gls{TD-SM} is given in \figref{fig:TDSM}. First, the incoming user bits are partitioned into $(k_\textrm{c}+k_\textrm{s})$-bits similar to the \gls{FD-SM}. Then, $k_\textrm{c}$ bits are used to modulate the $(\frac{N}{2}-1) \times 1$ subcarriers vector $\mathbf{q}$ by \gls{MQAM}. It should be noted that in \gls{TD-SM} technique, no spatial information exists in the frequency domain. Again, the Hermitian symmetry is imposed on the frequency domain frame as follows:
\begin{equation}
\mathbf{s}=[0~~\mathbf{q}^\textrm{T}~~0~~\tilde{\mathbf{q}}^\textrm{T}]^\textrm{T},
\end{equation}
\noindent where $\tilde{\mathbf{q}}$ is the conjugate symmetric vector of $\mathbf{q}$. Therefore, $s(k)=s(N-k)^\ast$ for $k\in \{ 0,1,\cdots,N-1 \}$. Due to the time domain spatial symbol encoding, a single \gls{OFDM} modulator/demodulator pair is always suffice in \gls{TD-SM}. Hence, the $N\times 1$ real and bipolar time domain samples vector $\mathbf{x}$ is obtained after the \gls{IFFT} operation by $\mathbf{x}=\mathbf{F}^{-1}\mathbf{s}$. In order to make the resultant signal unipolar, the vector $\mathbf{x}$ is \gls{DC}-biased and clipped similar to the \gls{FD-SM}. The \gls{PDF} for the elements of the biased and clipped vector $\bar{\mathbf{x}}$ before the spatial symbol encoding is the same as in \eqref{eq:pdftime}. In \gls{TD-SM}, only a single \gls{LED} is transmitting a non-zero value per time instant. The transmitting \gls{LED} is determined by the spatial mapping of $k_\textrm{s}$ bits as suggested by the \gls{SM} principle. To this end, in \gls{TD-SM} method, $N$ time domain samples are available to encode spatial symbols in contrast to \gls{FD-SM} which encodes $\frac{N}{2}-1$ subcarriers for that purpose. The \gls{PDF} for the elements in column vectors of the transmit matrix after the spatial symbol encoding is expressed by using \eqref{eq:pdftime} as follows:
\begin{equation}
f_{\bar{\mathbf{z}}_i}(v)=\frac{ f_{\bar{\mathbf{x}}}(v)(1-\delta(v))+(N_\textrm{t}-1)\delta(v)}{N_\textrm{t}}.
\label{eq:pdfTDSM}
\end{equation}
\noindent The $i^\textrm{th}$ column vector for the \gls{TD-SM} transmission matrix $\bar{\mathbf{Z}}$ is denoted by $\bar{\mathbf{z}}_i$. The \gls{PDF} of the biased and clipped transmit signal vector, $f_{\bar{\mathbf{x}}}(v)$, follows \eqref{eq:pdftime}. By using \eqref{eq:pdfTDSM}, the average transmitted electrical power per \gls{LED} in \gls{TD-SM} becomes
\begin{equation}
\textrm{E}\{ \bar{\mathbf{z}}^2_i \}=\frac{\textrm{E}\{ \bar{\mathbf{x}}^2 \}}{N_\textrm{t}}.
\label{eq:powerTDSM}
\end{equation}
\noindent where $\textrm{E}\{ \bar{\mathbf{x}}^2 \}$ can be calculated by using \eqref{eq:FDSMpower}. Thus, the total electrical transmit power of \gls{TD-SM} is also given by $P_{\textrm{e,TD-SM}}=\textrm{E}\{\bar{\mathbf{x}}^2\}$. It can be deduced from \eqref{eq:pdfTDSM} that even under the worst case scenario, where $L\neq 0 $, only a single \gls{LED} will be active per time in \gls{TD-SM} which implies \gls{ICI}-free transmission. As a result, noise and complexity enhancement emerging from the \gls{ZF} equalization at the receiver side are avoided in \gls{TD-SM}. It should also be noted that \gls{OWC} systems must satisfy the minimum illumination requirements along with communication functionality. Accordingly, the average illumination provided by the \gls{TD-SM} technique will be lower compared to \gls{FD-SM} due to $N_\textrm{t}-1$ idle \glspl{LED} in each time instant. However, this problem can easily be overcome by introducing a proper \gls{DC} bias to the idle \glspl{LED} without causing any performance degradation. Note that the illumination bias is intrinsically different than $B$ where $B$ is essential for signal transmission in \gls{IMDD}. Moreover, the illumination bias will not influence the effective \gls{SNR} of the system as it does not carry any information. Thus, the illumination bias is omitted in both simulations and experiments without loss of generality. Moving on, \gls{CP} added and digital-to-analog converted vectors are fed to the \glspl{LED} in a serial manner. Thus, the spectral efficiency of the \gls{TD-SM} becomes
\begin{equation}
\begin{split}
\eta_{\textrm{TD-SM}}&=\frac{1}{2}k_\textrm{c}G_\textrm{F}G_\textrm{T}+k_\textrm{s} \\ 
&\approx \frac{1}{2}k_\textrm{c}+k_\textrm{s}\quad \textrm{bpcu}.
\end{split}
\label{eq:TDSMspectral}
\end{equation}
\noindent It is obvious from \eqref{eq:FDSMspectral} and \eqref{eq:TDSMspectral} that the amount of data carried by the constellation symbols are the same in both \gls{FD-SM} and \gls{TD-SM}. However, due to the time-domain encoding of the spatial symbols, the amount of data conveyed in the spatial domain is doubled in the \gls{TD-SM} technique. The spectral efficiency comparisons between \gls{TD-SM}, \gls{FD-SM} and \gls{RC} are given for various numbers of transmitters, $N_\textrm{t}$, and various signal constellation sizes, $M$, in \figref{fig:secomparison}. In \gls{RC}, all the \glspl{LED} transmit the same signal \cite{fh1301}. It is observed from \figref{fig:secomparison} that \gls{TD-SM} achieves the highest spectral efficiency among the considered techniques for any number of \glspl{LED} and constellation orders.

At the RX side, the optical signal is captured and translated into the electrical domain by the \glspl{PD}. After analog-to-digital conversion and \gls{CP} removal, the real and positive valued baseband received electrical signal vector per time instant becomes
\begin{algorithm}[!t]
	\SetKwInOut{Input}{Input}
	\SetKwInOut{Output}{Output}
	\Input{$\mathbf{y}$, $\mathbf{H}$, $U$, $L$, $B$, $\sigma$ and $\sigma_\textrm{n}$.}
	\Output{Estimated time domain sample and spatial symbol vectors $\hat{\bar{\mathbf{x}}}=[\hat{\bar{x}}(0),\hat{\bar{x}}(1),\cdots,\hat{\bar{x}}(N-1)]^\textrm{T}$ and $\hat{\mathbf{i}}=[\hat{i}(0),\hat{i}(1),\cdots,\hat{i}(N-1)]^\textrm{T}$, respectively.}
	\textbf{Initialize}\;
	\For{t=0:(N-1)}{
	$\hat{i}(t)=\arg\min\limits_{1 \leq i\leq N_\textrm{t}}\left\{ \vert\vert \mathbf{y}-\mathbf{h}_i\hat{\bar{x}}^i(t) \vert \vert^2 + \frac{\sigma_\textrm{n}^2 \hat{\bar{x}}^i(t)}{\sigma^2}\left( \hat{\bar{x}}^i(t)-2B \right)\right\}$\;
	$\hat{\bar{x}}(t)=\hat{\bar{x}}^{\hat{i}(t)}(t)$\;
	}
	\caption{The algorithm for the joint \gls{MAP} estimator.}
	\label{algorithm}
\end{algorithm}
\begin{equation}
\mathbf{y}=\mathbf{h}_i\bar{x}(t)+\mathbf{n},
\label{eq:signalmodelTDSM}
\end{equation}
\noindent where $\mathbf{h}_i$ denotes the $i^\textrm{th}$ column vector of $\mathbf{H}$. The $t^\textrm{th}$ captured \gls{OFDM} time sample is given by $\bar{x}(t)$ for $t\in \{0,1\cdots,N-1\}$. The $N_\textrm{r}\times 1$ \gls{AWGN} vector is denoted by $\mathbf{n}$, which follows an i.i.d. Gaussian distribution, $\mathbf{n}\sim \mathcal{N}(0,\sigma_\textrm{n}^2)$. The spatial symbols in \gls{TD-SM} must be decoded in the time domain. According to \cite{bel1401}, the channel coupling effect on the spatial symbols is simply reversed by the \gls{ZF} equalizer. Please note that the performance degradation that emerges from the (near) rank deficient matrices in \gls{ZF} equalization of \gls{FD-SM} is completely avoided in \gls{TD-SM}. The effective channel is always rank one in \gls{TD-SM} as seen from \eqref{eq:signalmodelTDSM}. Therefore, the channel coupling can simply be reversed by the multiplication of reciprocals of the corresponding column vector elements. In other words, unlike \gls{FD-SM}, there is always a solution for \gls{ZF} equalization in \gls{TD-SM} as long as $H(j,i)\neq 0,~\forall i,j$. Due to the truncated Gaussian distributed characteristics of the time domain samples, the \gls{ZF} detector proposed in \cite{bel1401} becomes sub-optimal. In \cite{tyh1801}, the optimal \gls{MAP} estimator which jointly detects the transmitted samples and the \gls{LED} indexes is proposed. Note that the optimal \gls{MAP} estimator utilizes the \gls{PDF} of the time domain samples as well as the variance of the effective noise in the estimation process. The \gls{MAP} estimator used to detect the time domain samples is formulated by
\begin{figure}[!t]
	\centering
	\begin{subfigure}[t]{1\columnwidth}
		\centering
		\includegraphics[width=1\columnwidth,draft=false]{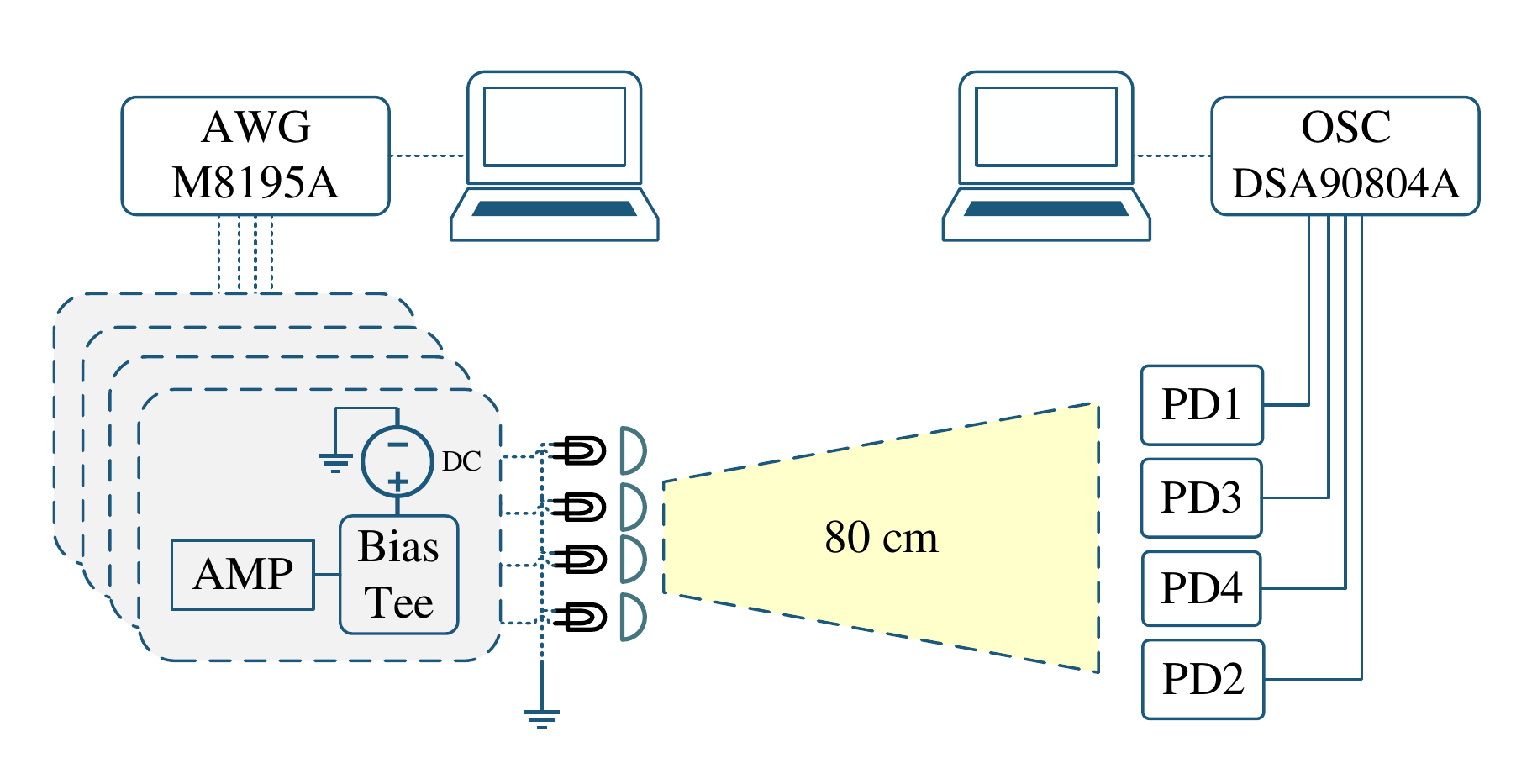}
		\caption{}
	\end{subfigure}\\
	\begin{subfigure}[t]{1\columnwidth}
		\centering
		\includegraphics[width=1\columnwidth,draft=false]{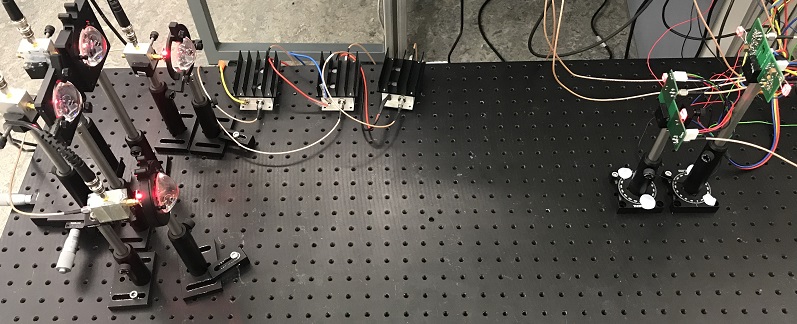}
		\caption{}
	\end{subfigure}
	\caption{The block diagram (a) and photo (b) of the experimental set-up.}
	\label{fig:expsetup}
\end{figure}
\begin{equation}
\hat{\bar{x}}(t)=\arg\max\limits_{\bar{x}(t)}p\left(\mathbf{y}~\vert~ \bar{x}(t)\right)p\left(\bar{x}(t)\right).
\label{eq:MAP1}\end{equation}
\noindent If we plug \eqref{eq:pdftime} into \eqref{eq:MAP1}, and after simple manipulations, the \gls{MAP} estimation of the transmitted sample becomes
\begin{equation}
\begin{split}
\Hat{\bar{x}}(t)&=\arg\min\limits_{\bar{x}(t)}\left\{ \mathcal{M}^\textrm{MAP} \right\}\\
&=\arg\min\limits_{\bar{x}(t)}\left\{ \sigma^2\norm{\mathbf{y}-\mathbf{h}_i\bar{x}(t)}^2+\sigma_\textrm{n}^2\left(\bar{x}(t)-B\right)^2 \right\}.
\end{split}
\label{eq:MAP2}\end{equation}
\noindent By taking the derivative of \eqref{eq:MAP2} and equating it to zero, we obtain the joint estimation of the $t^\textrm{th}$ sample value and $i^\textrm{th}$ active \gls{LED} index as follows:
\begin{equation}
\Hat{\bar{x}}^i(t)=\left[ \frac{\sigma^2(\mathbf{y}^\textrm{T}\mathbf{h}_i+\mathbf{h}_i^\textrm{T}\mathbf{y})+2B\sigma_\textrm{n}^2}{2(\sigma^2\mathbf{h}_i^\textrm{T}\mathbf{h}_i+\sigma_\textrm{n}^2)} \right]^U_L.
\label{eq:MAP3}\end{equation}
\noindent Note that \eqref{eq:MAP3} is conditioned on $i$. Hence, the active \gls{LED} index is estimated by the minimization of the \gls{MAP} metric, $\mathcal{M}^\textrm{MAP}$, given in \eqref{eq:MAP2}. After substituting $\bar{x}(t)$ in $\mathcal{M}^\textrm{MAP}$ by \eqref{eq:MAP3}, the estimated active \gls{LED} index becomes
\begin{equation}
\begin{split}
&\hat{i}(t)=\arg\min\limits_{1 \leq i\leq N_\textrm{t}}\left\{ \mathcal{M}^\textrm{MAP}\left( \Hat{\bar{x}}^i(t) \right) \right\}\\
&=\arg\min\limits_{1 \leq i\leq N_\textrm{t}}\left\{ \vert\vert \mathbf{y}-\mathbf{h}_i\hat{\bar{x}}^i(t) \vert \vert^2 + \frac{\sigma_\textrm{n}^2 \hat{\bar{x}}^i(t)}{\sigma^2}\left( \hat{\bar{x}}^i(t)-2B \right)\right\}.
\end{split}
\label{eq:MAPind}
\end{equation}
\noindent Then, the transmitted \gls{OFDM} sample value is determined by plugging \eqref{eq:MAPind} back into \eqref{eq:MAP3}. For further clarity, the joint \gls{MAP} estimation procedure is summarized in Algorithm \ref{algorithm}. Then, the \gls{FFT} of the estimated samples vector is taken by using $\hat{\mathbf{s}}=\mathbf{F}\hat{\bar{\mathbf{x}}}$. Finally, the transmitted user bits are recovered by demapping $\hat{\mathbf{s}}$ and $\hat{\mathbf{i}}$ jointly.

\section{Experimental Set-up and Methodology}\label{sec_setup}
In this section, the $4\times 4$ \gls{OFDM}-based optical \gls{SM} experimental set-up is explained along with the corresponding methodology for evaluating the performance of the system.

\setcounter{footnote}{0}

The block diagram and the photo of the experimental set-up are shown in Figs. \ref{fig:expsetup}(a) and \ref{fig:expsetup}(b), respectively. In the setup, $N_\textrm{t}=4$, VLMS1500-GS08 red \glspl{LED} are used where a Thorlabs ACL4532 aspheric condenser lens is placed in front of each \gls{LED} to collimate the output light of the \gls{LED}. Then, four digital signals are created in MATLAB using a laptop according to methods described in previous sections (i.e., \gls{TD-SM} or \gls{FD-SM}). In order to comply with the flat frequency response assumption, only 20 MHz of modulation bandwidth is considered. Four channels of a high speed \gls{AWG}, Keysight M8195A, is used to generate analog signals from the incoming digital samples. The sampling rate of the \gls{AWG} is 16 GSa/s and the resolution of the built-in \gls{DAC} unit is 8 bits. Signals generated by the \gls{AWG} are amplified by Mini-Circuits ZHL-1A-S+ amplifier modules (AMP). The \gls{DC} bias voltage and the forward current of \glspl{LED} are fixed at 2.3 V and 100 mA, respectively, which are controlled by a \gls{DC} power supply. These values are chosen to minimize the nonlinear effect of \glspl{LED}. The \gls{DC} bias is superimposed with the analog information carrying signal  through a bias-tee module, Mini-Circuits ZFBT-4R2GW and the resulting signal is fed to the \glspl{LED}. The link distance between the \gls{TX} and \gls{RX} is 80 cm.

\begin{figure}[!t]
\centering
\includegraphics[width=0.8\columnwidth]{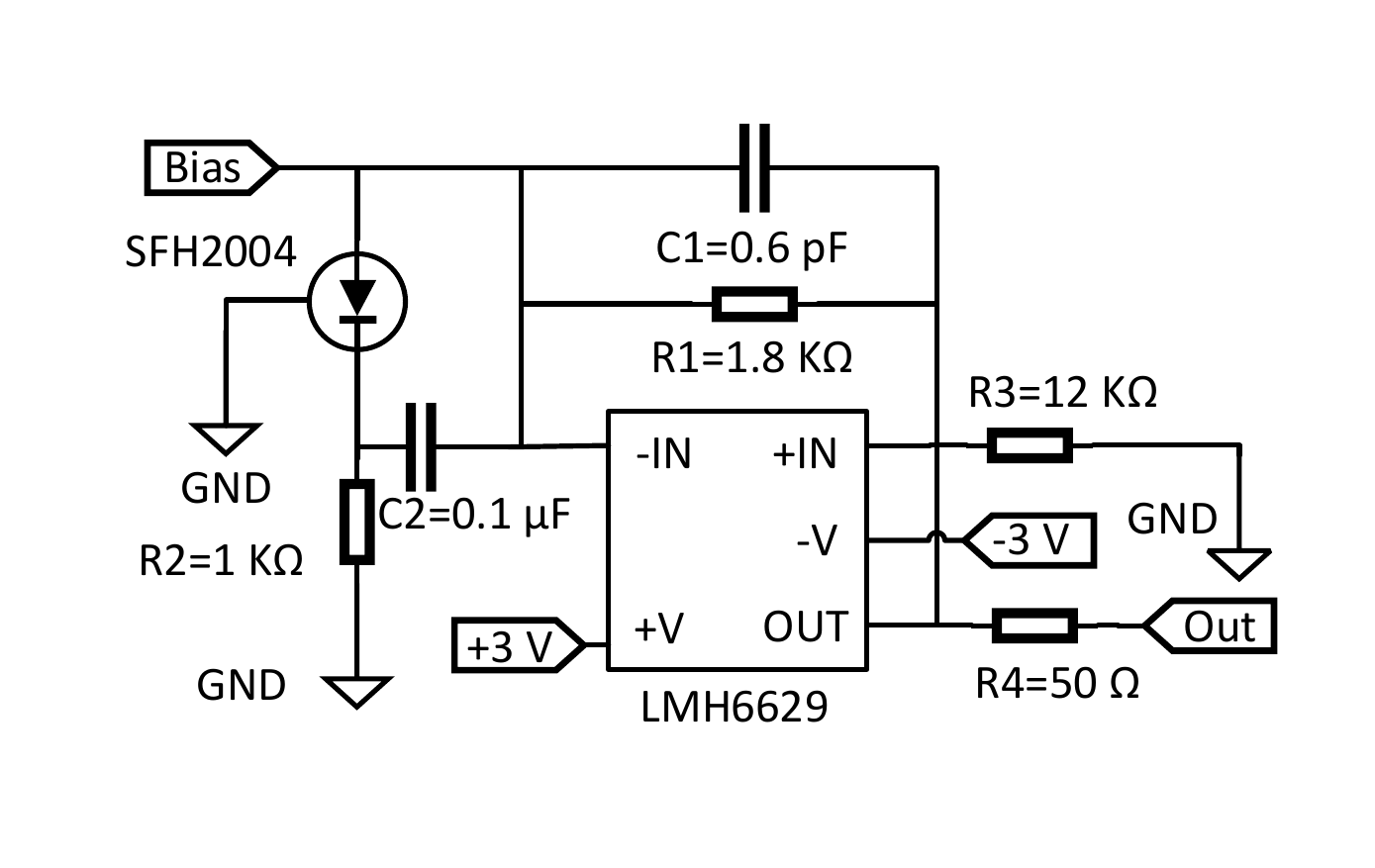}
\caption{Schematic presentation of PD circuit.}
\label{circuit}
\end{figure}

At the receiver, $N_\textrm{r}=4$, LEC-RP0508 high sensitivity positive-intrinsic-negative (PIN) \glspl{PD} are used in four receiver circuits as shown in Fig. \ref{circuit}. An operational amplifier chip, Texas Instruments LMH6629, is incorporated into the circuit. Subsequently, signals are captured by four channels of the Agilent DSA90804A high-speed oscilloscope (OSC). Lastly, the captured signals are sent to the laptop for post-processing in MATLAB.
\begin{figure}[!t]
\centering
\includegraphics[width=0.95\columnwidth]{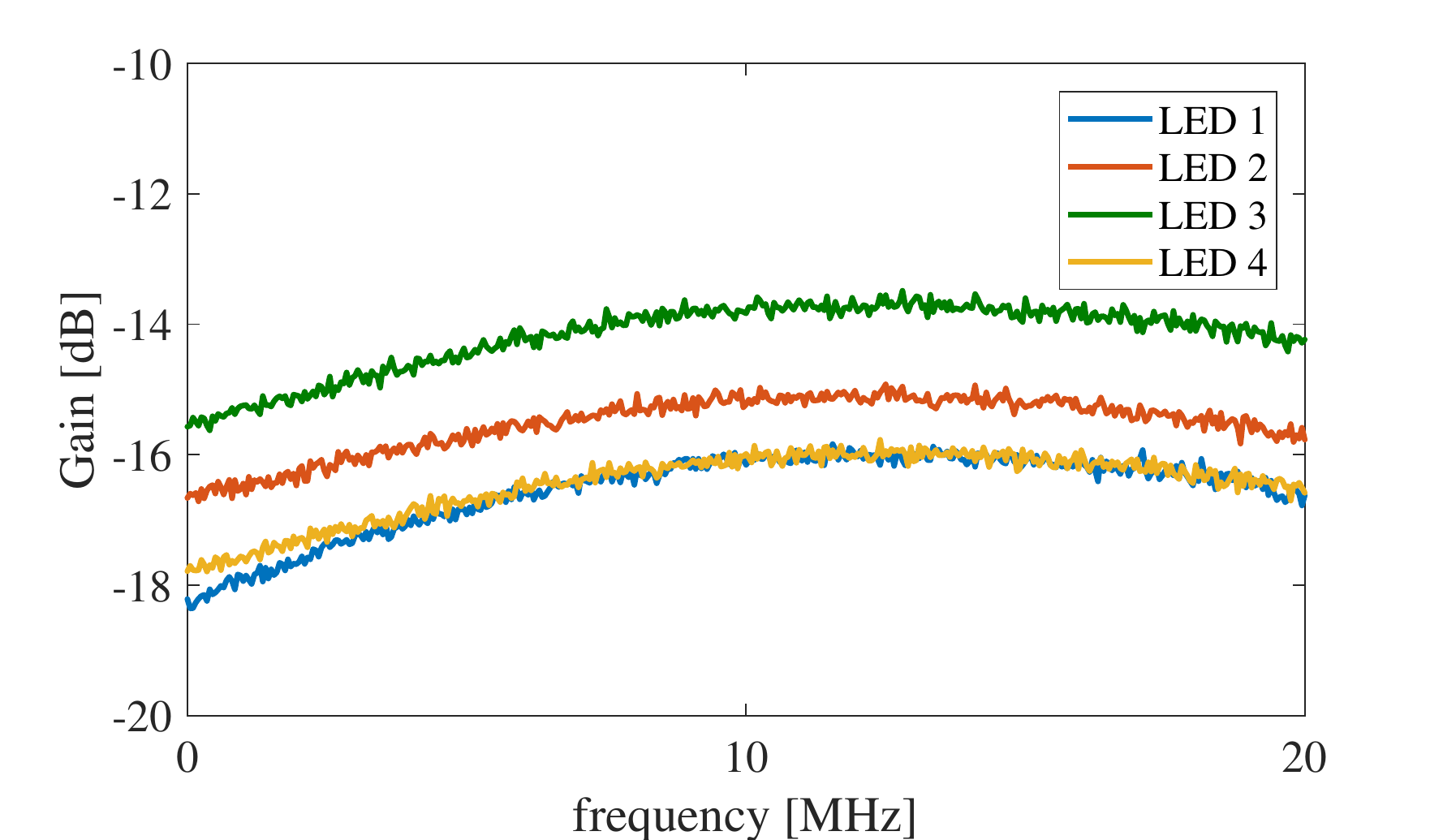}
\caption{Channel frequency responses of each LED for their dominant links.}
\label{freq_resp}
\end{figure}
\begin{figure}[!t]
\centering
\includegraphics[width=0.95\columnwidth]{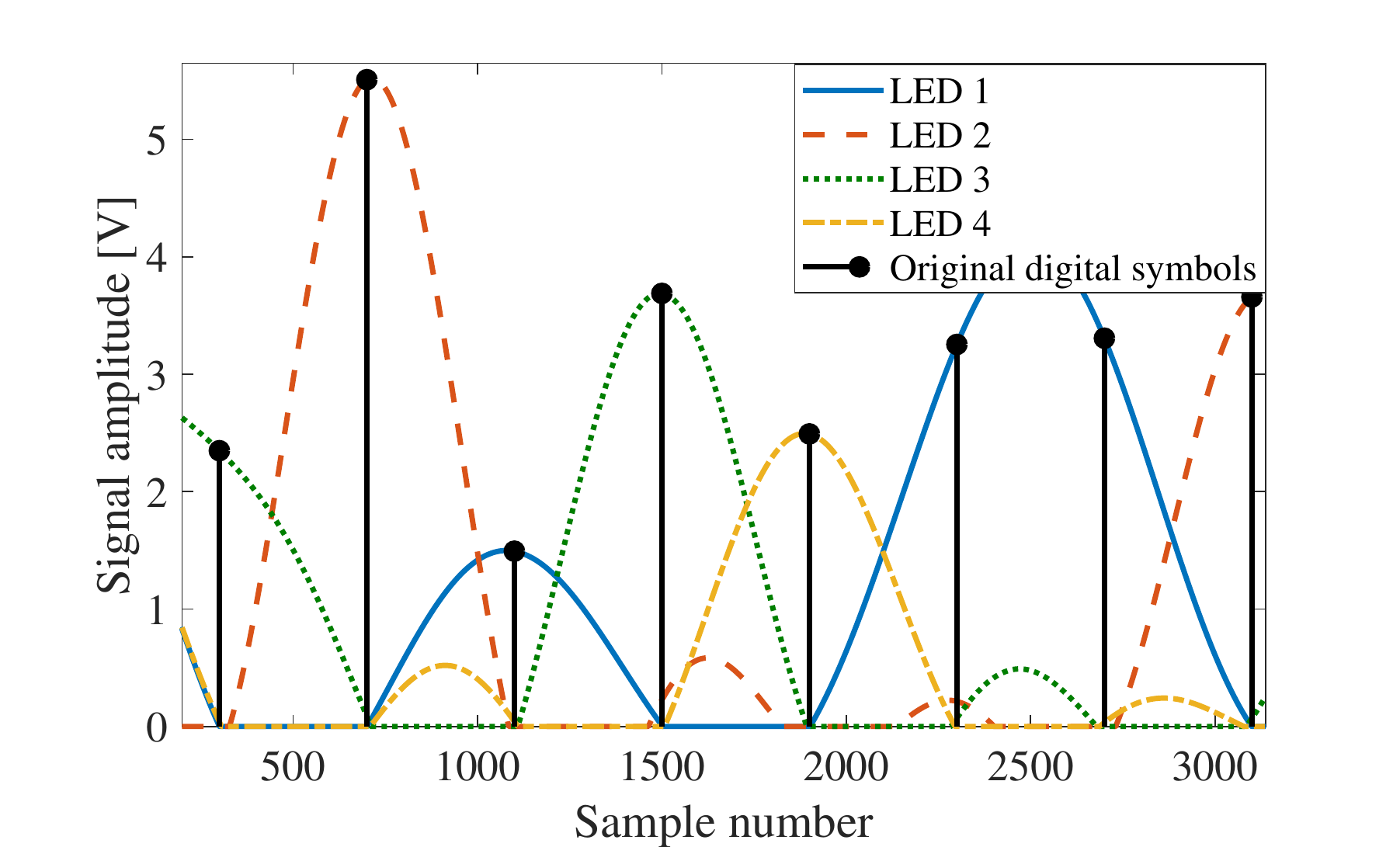}
\caption{An example frame of transmitted signals from different LEDs and the original digital signal for the TD-SM system.}
\label{example_TDSM}
\end{figure}

The channel matrix is estimated by training sequences, e.g., known OFDM symbols, prior to data transmission. Note that any error in the channel estimation may degrade the system performance. It is shown in the next section that simulation results match the experimental results, and thus, channel estimation in the assumed system model does not produce large errors. Furthermore, the likelihood of significant channel estimation errors is limited due to the quasi-static characteristics of the indoor optical wireless communications channel. The channel states remain stationary since the experimental setup does not include any moving components or other factors that may alter the channel condition. Therefore, an accurate initial channel estimation is valid throughout the signal transmission. The strongest front-end channel frequency responses, including the optical and electrical components, are shown in Fig. \ref{freq_resp} for each pair of transmitters and receivers. It is observed that frequency responses of the front-end system are almost flat within a 20 MHz modulation bandwidth, i.e., the frequency response stays within the 3~dB margin. The noise variance is estimated by comparing the oversampled received signal and the known transmitted signal. The estimated value is used for \gls{TD-SM} signal detection (see \eqref{eq:MAP3}). Part of the oversampled time domain signals and original \gls{OFDM} samples for \gls{TD-SM} are depicted in Fig. \ref{example_TDSM}. It can be seen in \figref{example_TDSM} that at specific time samples, which are designated for data transmission and detection, only one LED transmits a non-zero value. Obviously, synchronization of signals at the TX and RX is of vital importance. In this work, synchronization at the transmitter is ensured by using the single high accuracy \gls{AWG} with a built-in synchronization system. Moreover, at the beginning of each generated signal, a unique sequence of binary data is placed. The binary sequence is used to synchronize the received signals after they are captured by the oscilloscope.

As described in section \ref{sec_TDSM} and observed in Fig. \ref{example_TDSM}, an LED is active (i.e., ``ON") only when it is selected according to the spatial information. Thus, if an \gls{LED} is inactive (i.e., ``OFF"), it must be completely turned off, meaning that the \gls{DC} bias is also zero. However, as seen in Fig. \ref{fig:expsetup}(a), the \gls{DC} bias is provided by an external \gls{DC} power supply which cannot be controlled by the digital input signal. Therefore, the \gls{DC} bias is constantly applied to \glspl{LED}. This is different from the theoretical system model and the \gls{TD-SM} scheme. In order to solve this problem, the bipolar signal is modified before adding the \gls{DC} bias. For each \gls{LED}, zero values at the ``OFF'' time instants are replaced by a low negative value. Thus, the summation of the constant \gls{DC} bias and the bipolar signal yields a very small value at the corresponding time instants. Note that this modification is only necessary due to the hardware limitation in our experimental setup, and it could be rectified in future practical implementations of the \gls{TD-SM} system, where the DC bias and the signal are generated and controlled at the same time.

\begin{figure}[!t]
\centering
\includegraphics[width=1\columnwidth]{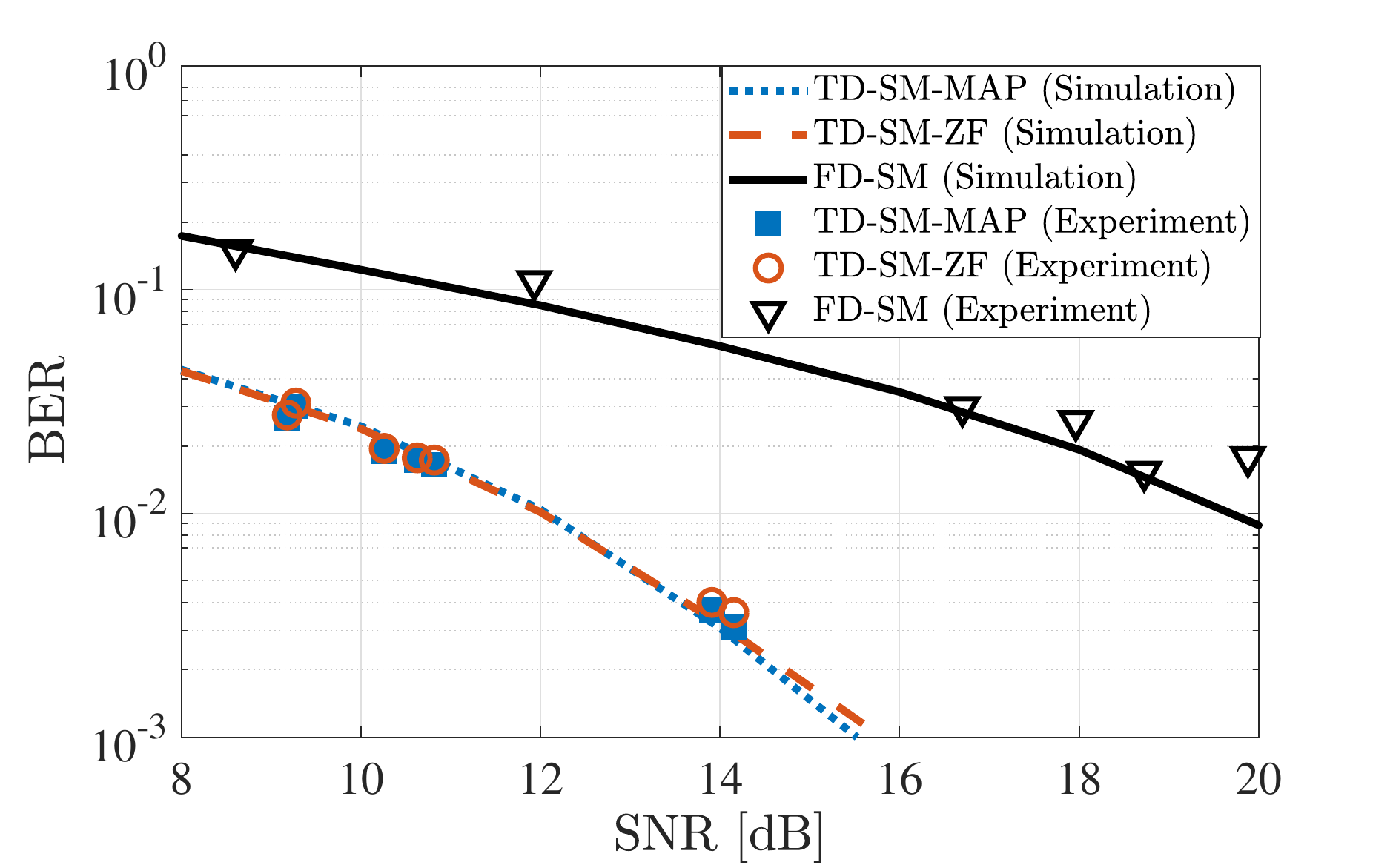}
\caption{BER for diagonal channel $\mathbf{H}_1$.}
\label{BER1}
\end{figure}
\section{Results and discussion}
In this section, the \gls{BER} performance of the described methods are compared. In order to ensure a fair comparison, the spectral efficiency of all methods is fixed at $\eta_{\textrm{FD-SM}}=\eta_{\textrm{TD-SM}}=4$ bpcu. Thus, 16-\gls{QAM} and 64-\gls{QAM} modulation formats are considered for \gls{TD-SM} and \gls{FD-SM}, respectively. The \gls{SNR} is calculated as the ratio between the total received electrical power and the noise at the \gls{RX}. Monte-Carlo simulation results are also presented. The optical channel gains and the noise variance obtained from the experimental results are used for Monte-Carlo simulations. The number of subcarriers and the \gls{DC} bias are chosen as $N=256$ and $B_\mathrm{dB}=10$ dB, respectively, for both the experimental system and  the computer simulations.

Firstly, the optical lenses in front of the \glspl{LED} are adjusted such that a diagonal channel matrix is realized. In other words, there is no overlap between optical links. The condition number of the channel matrix, $\rho$ defined as the ratio of the maximum eigenvalue to the minimum eigenvalue of the matrix, is equal to 1 in this case. The corresponding channel matrix is referred to as $\mathbf{H}_1$. Secondly, optical lenses are moved slightly in order to make the light spots larger at the receiver side so that they overlap with each other. Consequently, another channel matrix, namely $\mathbf{H}_2$, is obtained with a condition number greater than one, and in this case  $\rho=3.5$. In that scenario, almost 5\% of the transmitted optical power from each \gls{LED} is detected by two other \glspl{PD} in addition to the main \gls{PD}.
\begin{figure}[!t]
\centering
\includegraphics[width=1\columnwidth]{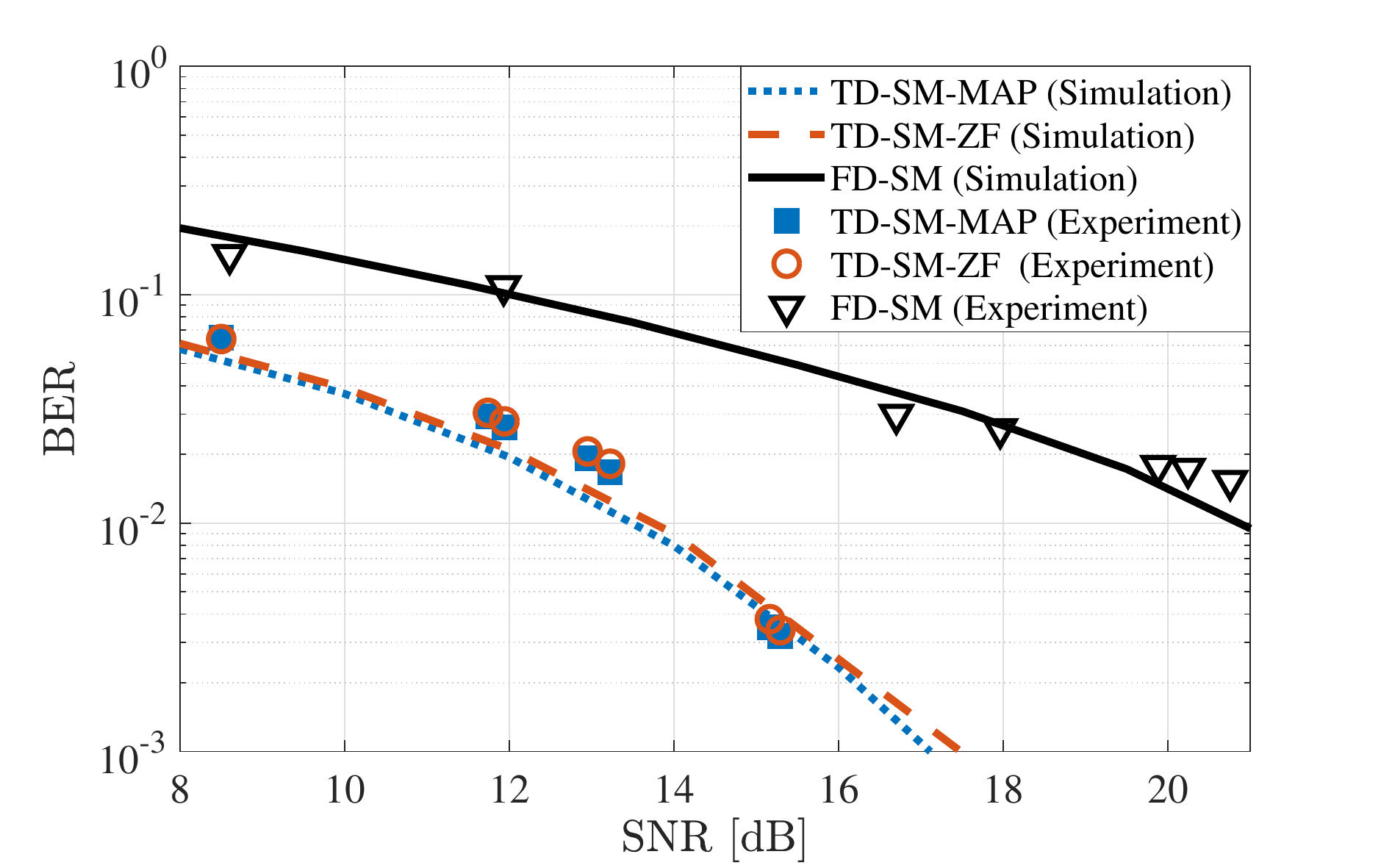}
\caption{BER for non-diagonal channel $\mathbf{H}_2$ with condition number $\rho=3.5$.}
\label{BER2}
\end{figure}

The \gls{BER} performance results are presented in Figs. \ref{BER1} and \ref{BER2}. Different \gls{SNR} values are obtained by scaling the digital signal at the \gls{AWG} prior to transmission. It is observed that in both scenarios, \gls{TD-SM} with either \gls{MAP} (TD-SM-MAP) or \gls{ZF} (TD-SM-ZF) detection outperforms \gls{FD-SM} by about 7 dB. It is also shown that the simulation results closely follow the experimental BER measurement for both \gls{FD-SM} and \gls{TD-SM}. Note that the same measured channel matrix from the experiment is used for simulations. In both Figs. \ref{BER1} and \ref{BER2}, the \gls{BER} of TD-SM-MAP is slightly lower than that of TD-SM-ZF. It is worth noting that higher SNR values would not be achievable in the experimental system due to the limited output optical power of the \glspl{LED}. In order to achieve high \gls{SNR} regime, the peak-to-peak amplitude of the transmit signal has to be increased which in turn worsens the effect of nonlinearity. The nonlinearity introduced by the front-end opto-electronic devices can simply be overcome by pre-distortion of the transmit signal\cite{dh1301}. For the sake of brevity, both \gls{FD-SM} and \gls{TD-SM} systems are presented with their simplest form without loss of generality.

It is important to emphasize that the difference between the \gls{FD-SM} and \gls{TD-SM} systems would be more significant in scenarios where the channel is ill-conditioned. As mentioned in the Section IV-B and presented in [21], \gls{TD-SM} exhibits inherent resilience against ill conditioned channels compared to \gls{FD-SM} due to a single active \gls{LED} per time instant. Thus, the effective channel in TD-SM is a column vector which is a rank one system. However, due to physical limitations in the experimental setup, we were not able to realize channel matrices with condition numbers higher than 3.5. Moreover, a larger overlap between different links (i.e., a larger light spot at the RX) significantly decreased the received optical power, and consequently, all system structures could only be realized in their low \gls{SNR} regime. Hence, the experimental demonstration of such scenarios is the subject of future work. In [53], it is shown that a larger array of transmitters and receivers may also change the channel condition number. In order to investigate this case, simulation results for a 16$\times$16 system are presented in Fig. 13 with $\eta_{\textrm{FD-SM}}=\eta_{\textrm{TD-SM}}=5$ bpcu and two different channel condition numbers of $\rho=1$ and $\rho=400$. To achieve the same spectral efficiency, 64-QAM and 4-QAM modulation formats are employed in \gls{FD-SM} and \gls{TD-SM}, respectively. It is observed that TD-SM-MAP achieves the lowest \gls{BER} in both channel conditions. For the $\rho=1$ case, \gls{TD-SM} with either \gls{MAP} or \gls{ZF} results in significantly lower BER compared to \gls{FD-SM} due to the lower modulation order $M$. A different behavior is observed for the $\rho=400$ case. While both \gls{FD-SM} and TD-SM-ZF fail, TD-SM-MAP demonstrates better performance because it benefits from \gls{MAP} detection.

In future work, the effect of temporal dispersion and intersymbol interference (i.e., effectively a frequency selective channel) on both FD-SM and TD-SM will be investigated. As the spatial symbols only exist in the time domain in TD-SM, a joint time-frequency domain detector will also be investigated to resolve the issue.

\begin{figure}[!t]
\centering
\includegraphics[width=1\columnwidth]{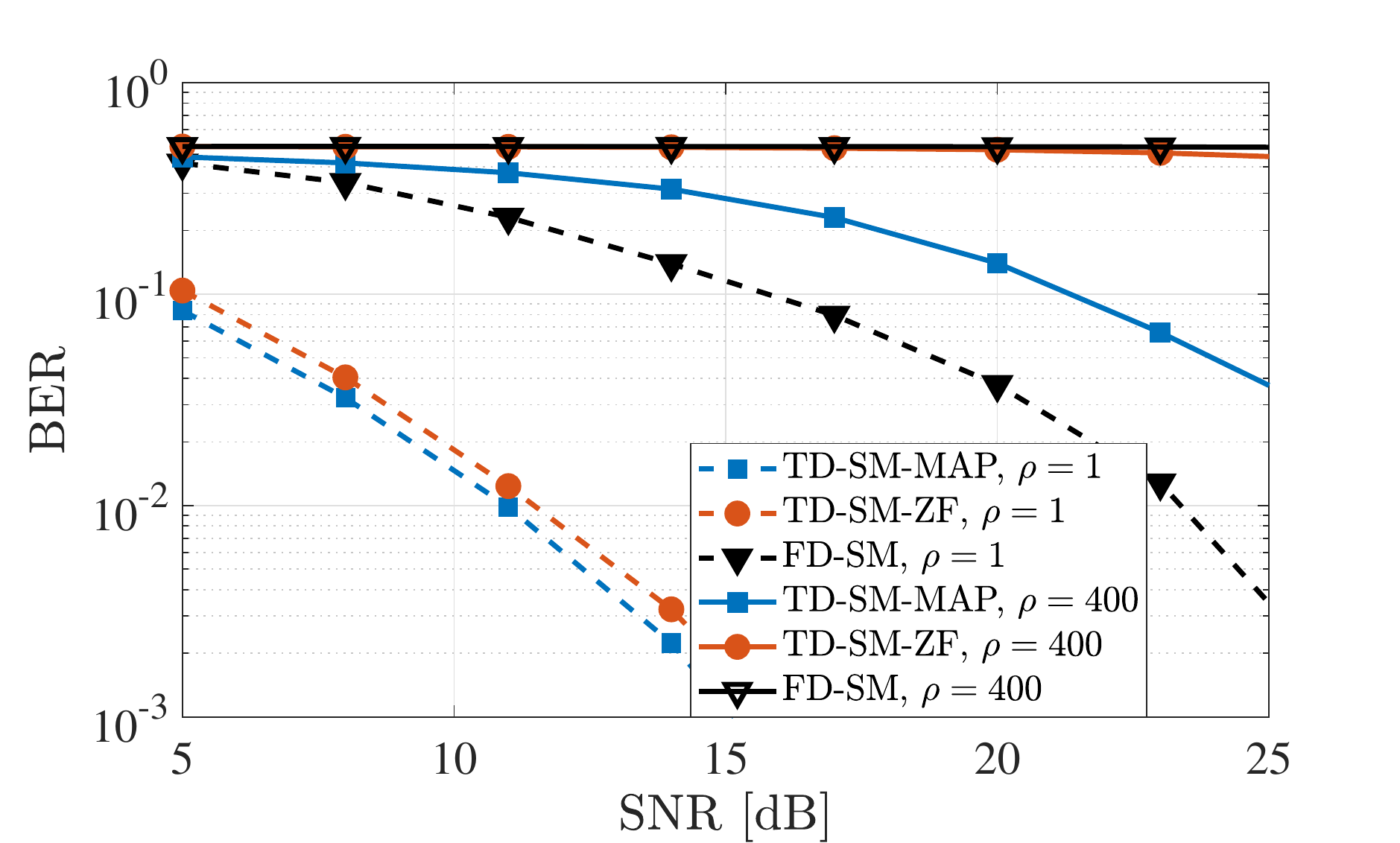}
\caption{BER results for a 16$\times$16 system with $\eta_{\textrm{FD-SM}}=\eta_{\textrm{TD-SM}}=5$ bpcu and channel condition numbers $\rho=1$ and $\rho=400$.}
\label{BER16}
\end{figure}



\section{Conclusion}
In this paper, two of the major \gls{OFDM}-based optical \gls{SM} techniques, namely \gls{FD-SM} and \gls{TD-SM} were investigated. The spatial symbols are encoded in the subcarrier level in \gls{FD-SM} which brings \gls{ICI}, \gls{TX}/\gls{RX} complexity and spectral efficiency loss. Conversely, in \gls{TD-SM} technique, the spatial symbols are encoded in the time domain as suggested by conventional \gls{SM}. It has been shown that \gls{TD-SM} inherits all the merits of conventional \gls{SM}. Moreover, \gls{TD-SM} techniques achieve significantly higher spectral efficiency compared to \gls{FD-SM}. Both systems have been validated  in terms of BER performance by Monte Carlo computer simulations as well as, for the first time ever, experimental results. It has also been shown by using extensive simulations and demonstrating the proof-of-concept experimental results that the \gls{TD-SM} with an optimal \gls{MAP} detector outperforms \gls{FD-SM} in terms of the \gls{BER} performance. Consequently, the experimental results verify and suggest that \gls{TD-SM} is a viable candidate for next-generation \gls{OFDM}-based \gls{SM}. 




\section*{Acknowledgment}
Professor Haas gratefully acknowledge the support of this research by the Engineering and Physical Sciences Research Council (EPSRC) under an Established Career Fellowship grant, EP/R007101/1. He also acknowledges the financial support of his research by the Wolfson Foundation and the Royal Society. Anil Yesilkaya acknowledges financial support for his PhD studies from Zodiac Inflight Innovations (TriaGnoSys GmbH). The authors would like to thank Dr. Tezcan Cogalan and Dr. Hossein Kazemi for their insightful comments and suggestions.

\bibliography{IEEEabrv,kitap2018}
\bibliographystyle{IEEEtran}

\begin{IEEEbiography}[{\includegraphics[width=1in,height=1.25in,clip,keepaspectratio]{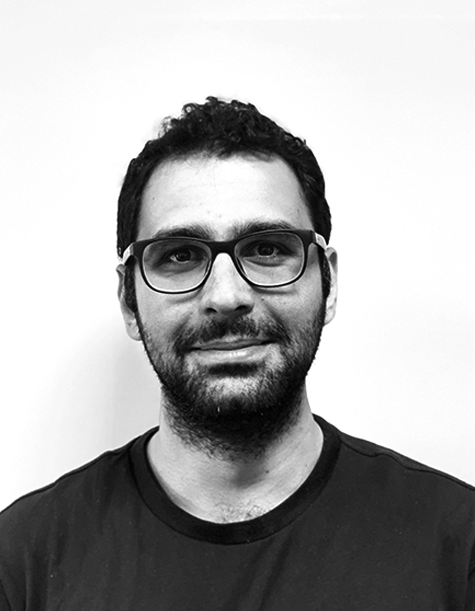}}]{Anil Yesilkaya}
(S’11) received the B.Sc (1st Hons) and M.Sc degrees in electronics engineering from the Kadir Has University, Istanbul, Turkey in 2014 and 2016, respectively. He was the recipient of the Best Paper Award at the IEEE International Conference on Communications, in 2018. He is currently pursuing the Ph.D. degree in digital communications with the University of Edinburgh. His research interests include, PHY layer security, multiple-input multiple-output optical wireless communications and LiFi-based in-flight connectivity.
\end{IEEEbiography}

\begin{IEEEbiography}[{\includegraphics[width=1in,height=1.25in,clip,keepaspectratio]{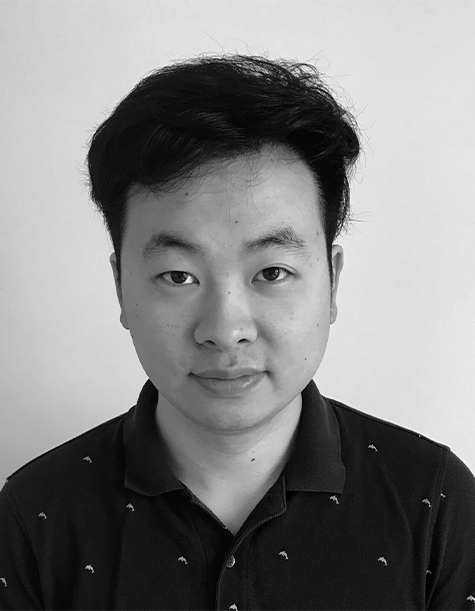}}]{Rui Bian}
received the B.Eng. degree in information engineering from the Wuhan University of Technology, Wuhan, China, in 2011, and the M.Sc. degree in electronics from the University of Edinburgh, Edinburgh, U.K., in 2013, where he is currently pursuing the Ph.D. degree in digital communications. His main research interest is in visible light communication.
\end{IEEEbiography}

\begin{IEEEbiography}[{\includegraphics[width=1in,height=1.25in,clip,keepaspectratio]{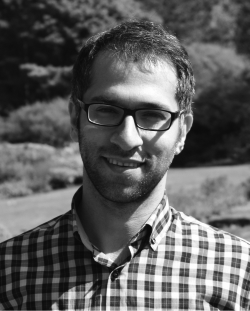}}]{Iman Tavakkolnia}
(S’15–M’18) received the B.Sc. degree in telecommunication engineering from the University of Tehran, Tehran, Iran, in 2006, the M.Sc. degree in communication systems from the Sharif University of Technology, Tehran, Iran, in 2011, and the Ph.D. degree in electrical engineering from the University of Edinburgh, Edinburgh, U.K., in 2018. He is currently a Research Associate with LiFi Research and Development Centre, University of Edinburgh. His research interests include communication theory, optical fiber communication, and visible light communication.
\end{IEEEbiography}

\begin{IEEEbiography}[{\includegraphics[width=1in,height=1.25in,clip,keepaspectratio]{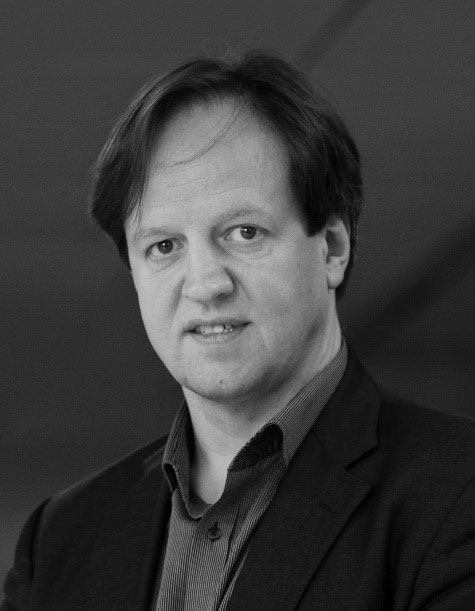}}]{Harald Haas}
(S’98-AM’00-M’03-SM’16-F’17) received the Ph.D. degree from The University of Edinburgh in 2001. He is currently the Chair of Mobile Communications at The University of Edinburgh, and he is the Initiator, Co-Founder, and Chief Scientific Officer of pureLiFi Ltd., and the Director of the LiFi Research and Development Centre, The University of Edinburgh. He has authored 430 conference and journal papers, including a paper in Science and co-authored a book Principles of LED Light Communications Towards Networked Li-Fi (Cambridge University Press, 2015). His main research interests are in optical wireless communications, hybrid optical wireless and RF communications, spatial modulation, and interference coordination in wireless networks. He first introduced and coined spatial modulation and LiFi. LiFi was listed among the 50 best inventions in TIME Magazine in 2011. He was an invited speaker at TED Global 2011, and his talk on Wireless Data from Every Light Bulb has been watched online over 2.4 million times. He gave a second TED Global lecture in 2015 on the use of solar cells as LiFi data detectors and energy harvesters. This has been viewed online over 2 million times. He was elected as a fellow of the Royal Society of Edinburgh and a Fellow of the IEEE in 2017. In 2012 and 2017, he was a recipient of the prestigious Established Career Fellowship from the Engineering and Physical Sciences Research Council (EPSRC) within Information and Communications Technology in the U.K. In 2014, he was selected by EPSRC as one of ten Recognizing Inspirational Scientists and Engineers Leaders in the U.K. He was a corecipient of the EURASIP Best Paper Award for the Journal on Wireless Communications and Networking in 2015 and the Jack Neubauer Memorial Award of the IEEE Vehicular Technology Society. In 2016, he received the Outstanding Achievement Award from the International Solid State Lighting Alliance. He was a co-recipient of recent best paper awards at VTC-Fall, 2013, VTC-Spring 2015, ICC 2016, ICC 2017 and ICC 2018. He is an Editor of the IEEE TRANSACTIONS ON COMMUNICATIONS and the IEEE JOURNAL OF LIGHTWAVE TECHNOLOGIES.
\end{IEEEbiography}
\end{document}